\documentclass{article}

 \usepackage[preprint]{neurips_2026}

\usepackage[utf8]{inputenc} 
\usepackage[T1]{fontenc}    
\usepackage{hyperref}       
\usepackage{url}            
\usepackage{booktabs}       
\usepackage{amsfonts}       
\usepackage{nicefrac}       
\usepackage{microtype}      
\usepackage{xcolor}         
\usepackage{amsfonts}       
\usepackage{nicefrac}       
\usepackage{microtype}      
\usepackage{xcolor}         
\usepackage{colortbl, booktabs}
\usepackage{natbib}
\usepackage{graphicx}
\usepackage{array}
\usepackage{caption}
\usepackage{algorithm}
\usepackage{algorithmic}
\usepackage{arydshln}
\usepackage{wrapfig}
\usepackage{subcaption}
\usepackage{float}
\usepackage{siunitx}
\usepackage{color}
\usepackage{makecell}
\usepackage{bbding} 
\usepackage{amssymb}
\usepackage{amsmath}
\usepackage{multirow}
\usepackage{bm}
\usepackage{wrapfig} 
\usepackage[most]{tcolorbox}
\usepackage[table]{xcolor}
\definecolor{brainagentgreen}{RGB}{232, 246, 232}

\definecolor{thinkcolor}{RGB}{227,196,144}
\definecolor{observecolor}{RGB}{153,201,227}
\definecolor{explorecolor}{RGB}{178,217,200}
\definecolor{interest_colframe}{rgb}{0.8, 0.878, 0.871}
\definecolor{interest_colback}{rgb}{0.918, 0.953, 0.949}

\definecolor{tag_colframe}{rgb}{0.965, 0.898, 0.847}
\definecolor{tag_colback}{rgb}{0.988, 0.961, 0.941}

\definecolor{exp_colback}{rgb}{0.949, 0.965, 0.980}
\definecolor{exp_colframe}{rgb}{0.878, 0.922, 0.965}

\newtcolorbox{promptbox}[1]{
    enhanced,
    width=\textwidth,        
    colback=white,
    colframe=gray!50,
    boxrule=0.5pt,
    toprule=2.5pt, 
    coltitle=black,
    fonttitle=\bfseries\sffamily,
    title={#1},
    sharp corners, 
    left=8pt, right=8pt, top=8pt, bottom=8pt,
    after skip=1.5em,
    boxed title style={empty, size=minimal, top=2pt, bottom=5pt},
}

\newtcolorbox{inferbox}[1]{
    enhanced,
    width=\textwidth,        
    colback=white,
    colframe=interest_colframe,
    boxrule=0.5pt,
    toprule=2.5pt, 
    coltitle=black,
    fonttitle=\bfseries\sffamily,
    title={#1},
    sharp corners, 
    left=8pt, right=8pt, top=8pt, bottom=8pt,
    after skip=1.5em,
    boxed title style={empty, size=minimal, top=2pt, bottom=5pt},
}

\usepackage{listings}

\definecolor{brainagentTitle}{RGB}{40,96,160}
\definecolor{brainagentBack}{RGB}{246,249,253}
\definecolor{brainagentFrame}{RGB}{120,150,190}
\definecolor{brainagentText}{RGB}{35,45,60}

\lstdefinelanguage{json}{
    basicstyle=\scriptsize\ttfamily\color{brainagentText},
    showstringspaces=false,
    breaklines=true,
    columns=fullflexible,
    morestring=[b]",
    stringstyle=\color{brainagentText},
    literate=
     *{0}{{{\color{brainagentText}0}}}{1}
      {1}{{{\color{brainagentText}1}}}{1}
      {2}{{{\color{brainagentText}2}}}{1}
      {3}{{{\color{brainagentText}3}}}{1}
      {4}{{{\color{brainagentText}4}}}{1}
      {5}{{{\color{brainagentText}5}}}{1}
      {6}{{{\color{brainagentText}6}}}{1}
      {7}{{{\color{brainagentText}7}}}{1}
      {8}{{{\color{brainagentText}8}}}{1}
      {9}{{{\color{brainagentText}9}}}{1}
}

\newtcblisting{brainagentoutput}{
    enhanced,
    breakable,
    listing only,
    colback=brainagentBack,
    colframe=brainagentFrame,
    colbacktitle=brainagentTitle,
    coltitle=white,
    title={Structured Output of BrainAgent},
    fonttitle=\bfseries\small,
    boxrule=0.8pt,
    arc=1mm,
    left=1mm,
    right=1mm,
    top=1mm,
    bottom=1mm,
    listing options={
        language=json,
        basicstyle=\scriptsize\ttfamily\color{brainagentText},
        breaklines=true,
        showstringspaces=false,
        columns=fullflexible
    }
}

\definecolor{brainagentPurple}{RGB}{170,30,120}
\definecolor{brainagentBack}{RGB}{252,247,253}
\definecolor{jsonGreen}{RGB}{75,150,55}
\definecolor{jsonBlue}{RGB}{70,90,170}

\lstdefinelanguage{json}{
    basicstyle=\tiny\ttfamily\color{jsonGreen},
    showstringspaces=false,
    breaklines=true,
    columns=fullflexible,
    morestring=[b]",
    stringstyle=\color{jsonGreen}
}

\newtcblisting{reasoningbox}{
    enhanced,
    listing only,
    breakable,
    colback=brainagentBack,
    colframe=black,
    colbacktitle=brainagentPurple,
    coltitle=white,
    title={Understanding-stage Reasoning Result},
    fonttitle=\bfseries\large,
    boxrule=0.9pt,
    arc=1mm,
    left=1.5mm,
    right=1.5mm,
    top=1mm,
    bottom=1mm,
    listing options={
        language=json,
        basicstyle=\scriptsize\ttfamily\color{jsonGreen},
        breaklines=true,
        showstringspaces=false,
        columns=fullflexible
    }
}

\definecolor{thinkTitle}{RGB}{105,125,115}
\definecolor{actionTitle}{RGB}{170,30,120}
\definecolor{boxBack}{RGB}{252,248,253}
\definecolor{jsonGreen}{RGB}{75,150,55}

\newtcblisting{thinkbox}{
    enhanced,
    listing only,
    breakable,
    colback=boxBack,
    colframe=black,
    colbacktitle=thinkTitle,
    coltitle=white,
    title={Report-stage Reasoning Result (Round 1)},
    fonttitle=\bfseries\small,
    boxrule=0.9pt,
    arc=1mm,
    left=1mm,
    right=1mm,
    top=1mm,
    bottom=1mm,
    listing options={
        language=json,
        basicstyle=\scriptsize\ttfamily\color{jsonGreen},
        breaklines=true,
        showstringspaces=false,
        columns=fullflexible
    }
}

\newtcblisting{actionbox}{
    enhanced,
    listing only,
    breakable,
    colback=boxBack,
    colframe=black,
    colbacktitle=actionTitle,
    coltitle=white,
    title={Action-stage Reasoning Result (Round 1)},
    fonttitle=\bfseries\small,
    boxrule=0.9pt,
    arc=1mm,
    left=1mm,
    right=1mm,
    top=1mm,
    bottom=1mm,
    listing options={
        language=json,
        basicstyle=\tiny\ttfamily\color{jsonGreen},
        breaklines=true,
        showstringspaces=false,
        columns=fullflexible
    }
}

\newtcblisting{thinkboxroundtwo}{
    enhanced,
    listing only,
    breakable,
    colback=boxBack,
    colframe=black,
    colbacktitle=thinkTitle,
    coltitle=white,
    title={Report-stage Reasoning Result (Round 2)},
    fonttitle=\bfseries\small,
    boxrule=0.9pt,
    arc=1mm,
    left=1mm,
    right=1mm,
    top=1mm,
    bottom=1mm,
    width=\textwidth,
    listing options={
        language=json,
        basicstyle=\tiny\ttfamily\color{jsonGreen},
        breaklines=true,
        showstringspaces=false,
        columns=fullflexible
    }
}

\newtcblisting{actionboxroundtwo}{
    enhanced,
    listing only,
    breakable,
    colback=boxBack,
    colframe=black,
    colbacktitle=actionTitle,
    coltitle=white,
    title={Action-stage Reasoning Result (Round 2)},
    fonttitle=\bfseries\small,
    boxrule=0.9pt,
    arc=1mm,
    left=1mm,
    right=1mm,
    top=1mm,
    bottom=1mm,
    width=\textwidth,
    listing options={
        language=json,
        basicstyle=\tiny\ttfamily\color{jsonGreen},
        breaklines=true,
        showstringspaces=false,
        columns=fullflexible
    }
}

\newtcblisting{thinkboxroundthree}{
    enhanced,
    listing only,
    breakable,
    colback=boxBack,
    colframe=black,
    colbacktitle=thinkTitle,
    coltitle=white,
    title={Report-stage Reasoning Result (Round 3)},
    fonttitle=\bfseries\small,
    boxrule=0.9pt,
    arc=1mm,
    left=1mm,
    right=1mm,
    top=1mm,
    bottom=1mm,
    width=\textwidth,
    listing options={
        language=json,
        basicstyle=\tiny\ttfamily\color{jsonGreen},
        breaklines=true,
        showstringspaces=false,
        columns=fullflexible
    }
}

\newtcblisting{analysisboxroundthree}{
    enhanced,
    listing only,
    breakable,
    colback=boxBack,
    colframe=black,
    colbacktitle=actionTitle,
    coltitle=white,
    title={Final Analysis Reasoning Result (Round 3)},
    fonttitle=\bfseries\small,
    boxrule=0.9pt,
    arc=1mm,
    left=1mm,
    right=1mm,
    top=1mm,
    bottom=1mm,
    width=\textwidth,
    listing options={
        language=json,
        basicstyle=\tiny\ttfamily\color{jsonGreen},
        breaklines=true,
        showstringspaces=false,
        columns=fullflexible
    }
}

\newtcblisting{reflectionbox}{
    enhanced,
    listing only,
    breakable,
    colback=boxBack,
    colframe=black,
    colbacktitle=actionTitle,
    coltitle=white,
    title={Reflection-stage Reasoning Result},
    fonttitle=\bfseries\small,
    boxrule=0.9pt,
    arc=1mm,
    left=1mm,
    right=1mm,
    top=1mm,
    bottom=1mm,
    width=\textwidth,
    listing options={
        language=json,
        basicstyle=\tiny\ttfamily\color{jsonGreen},
        breaklines=true,
        showstringspaces=false,
        columns=fullflexible
    }
}

\definecolor{brainagentTitle}{RGB}{54,130,78}
\definecolor{directTitle}{RGB}{170,85,65}
\definecolor{jsonBack}{RGB}{248,250,248}
\definecolor{jsonText}{RGB}{40,70,45}

\lstdefinelanguage{json}{
    basicstyle=\scriptsize\ttfamily\color{jsonText},
    showstringspaces=false,
    breaklines=true,
    columns=fullflexible,
    morestring=[b]",
    stringstyle=\color{jsonText}
}

\newtcblisting{brainagentjsoncase}{
    enhanced,
    breakable,
    listing only,
    colback=jsonBack,
    colframe=brainagentTitle,
    colbacktitle=brainagentTitle,
    coltitle=white,
    title={Grok + BrainAgent},
    fonttitle=\bfseries\small,
    boxrule=0.8pt,
    arc=1mm,
    left=1mm,
    right=1mm,
    top=1mm,
    bottom=1mm,
    width=\textwidth,
    listing options={
        language=json,
        basicstyle=\scriptsize\ttfamily\color{jsonText},
        breaklines=true,
        showstringspaces=false,
        columns=fullflexible
    }
}

\newtcblisting{directjsoncase}{
    enhanced,
    breakable,
    listing only,
    colback=jsonBack,
    colframe=directTitle,
    colbacktitle=directTitle,
    coltitle=white,
    title={Direct Grok Inference},
    fonttitle=\bfseries\small,
    boxrule=0.8pt,
    arc=1mm,
    left=1mm,
    right=1mm,
    top=1mm,
    bottom=1mm,
    width=\textwidth,
    listing options={
        language=json,
        basicstyle=\scriptsize\ttfamily\color{jsonText},
        breaklines=true,
        showstringspaces=false,
        columns=fullflexible
    }
}

\title{%
  \raisebox{-0.3ex}{\includegraphics[width=0.8cm]{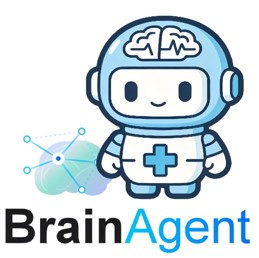}}%
  \hspace{3.5px}When Language Models Meet NeuroGraphs: Exploring Enhanced Agentic LLM Framework Towards Brain Network Analysis%
}

\author{%
  Jiaxing Li$^{1}$\thanks{Equal contribution. Code can be available at https://github.com/KamonRiderDR/BrainAgent.} \hspace{1.5em}
  Rui Dong$^{1*}$ \hspace{1.5em}
  Muyao Tang$^1$ \hspace{1.5em}
  Youyong Kong$^{1}$\thanks{Corresponding author.} \hspace{1.5em} \\
  \textsuperscript{1}School of Computer Science and Engineering, Southeast University \\
  \texttt{jiaxing\_li@seu.edu.cn, 1436598249@qq.com, mytang@seu.edu.cn} \\
  \texttt{kongyouyong@seu.edu.cn}
}

\begin{document}

\maketitle

\begin{abstract}
Brain network analysis is crucial for understanding cognition and neurological disorders, yet existing deep learning methods mainly treat connectome analysis as a graph-to-logit classification problem, offering limited explanatory reasoning. Large language models (LLMs) provide a promising interface for knowledge-intensive scientific analysis, but directly applying general-purpose LLMs to brain networks remains challenging due to the structure-language gap, limited neuroscience grounding, and overconfident positive predictions. In this paper, we propose \textbf{BrainAgent}, an agentic LLM framework for knowledge-enhanced brain network analysis. BrainAgent reformulates connectome classification as an iterative process of topology-aware understanding, external retrieval, reasoning, and reflection. Specifically, it first converts raw brain networks into compact multi-level structural descriptions through brain-specific analysis tools, then retrieves relevant neuroscience knowledge and task-specific cases to ground the reasoning process, and finally generates structured predictions with reflective verification. Experiments on four public rs-fMRI datasets show that BrainAgent consistently improves different closed-source and open-source LLM backbones over direct prompting and standard reasoning baselines. Further ablation and interpretability analyses demonstrate the effectiveness of each component and show that BrainAgent produces more comprehensive, multi-level, and verifiable explanations.These results indicate that agentic LLMs provide a practical route toward interpretable and knowledge-grounded brain network analysis.
\end{abstract}

\section{Introduction}
\label{sec:intro}

Brain network analysis provides a principled way to study human cognition and neurological disorders by representing neuroimaging measurements as graphs, where nodes denote brain regions and edges encode structural or functional interactions~\citep{bullmore2009complex}. In particular, resting-state functional MRI (rs-fMRI) enables the construction of functional connectomes and has been widely used for disease-related classification and biomarker discovery. However, brain network analysis is not merely a prediction problem. For clinical and scientific use, a model is expected to explain \emph{why} certain topological patterns are associated with a disorder-related label, and how such evidence is grounded in known neuroscience knowledge.

\begin{figure*}[t]
  \centering
  \includegraphics[width=\textwidth]{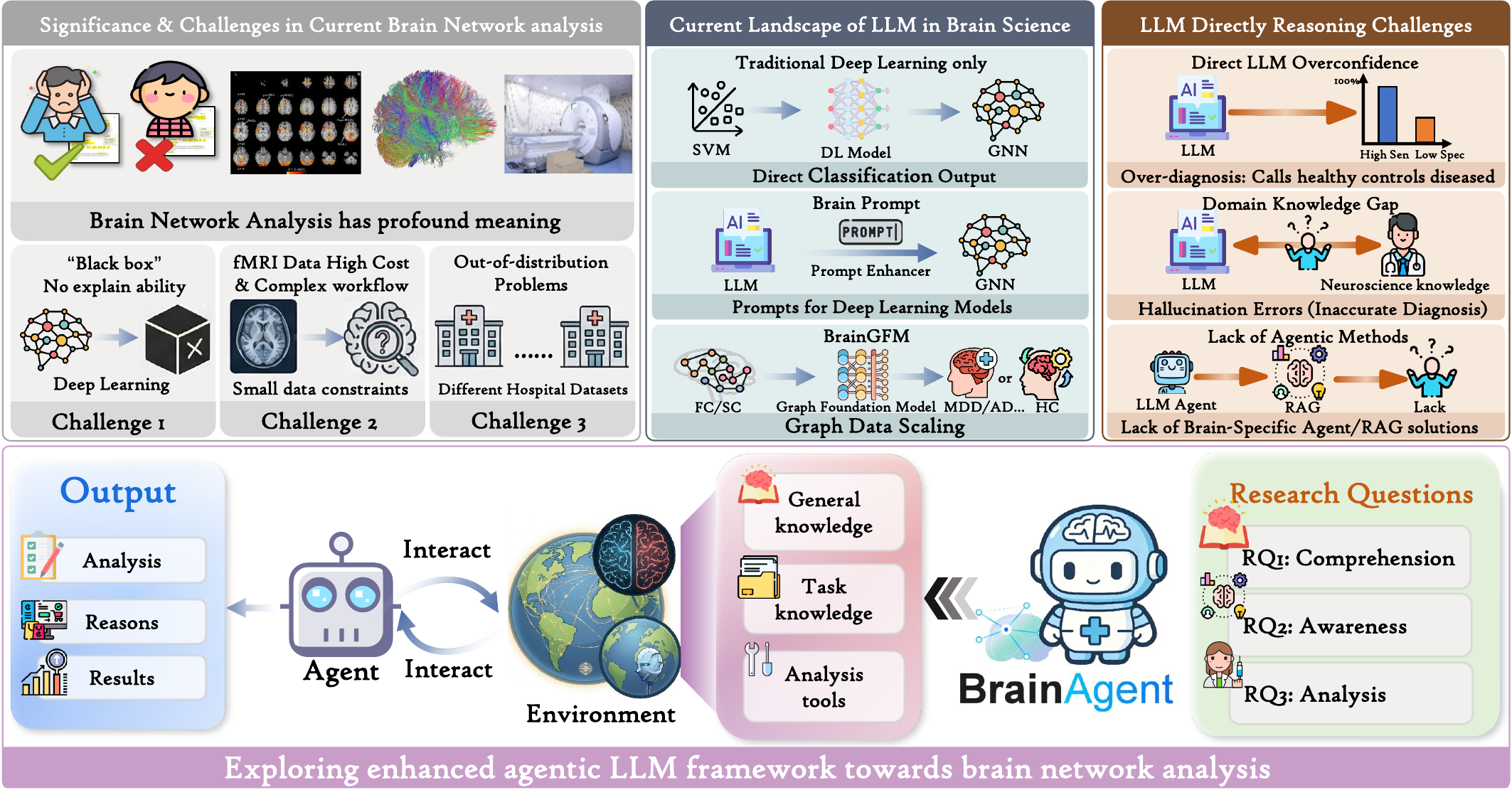}
  \caption{Illustration of the motivation and overall framework of \textbf{BrainAgent}. We position brain network analysis as a knowledge-intensive, agentic reasoning problem. BrainAgent addresses this problem through topology-aware graph understanding, tool-augmented interaction, neuroscience knowledge retrieval, task-specific case retrieval, and reflective analysis.}
  \label{fig:teaser}
\end{figure*}

Existing methods mainly formulate brain network analysis as a supervised graph classification task. Graph neural networks (GNNs)~\citep{gcn,gat,li2021braingnn,NeuroField} and standardized GNN benchmarks for brain networks~\citep{cui2023braingb} have achieved promising performance by learning discriminative graph representations. Nevertheless, this representation-centric paradigm has intrinsic limitations (Figure \ref{fig:teaser}). First, most models map a brain graph to logits through a fixed forward pass, providing limited explanatory reasoning beyond post-hoc saliency or region visualization. Second, rs-fMRI data are typically expensive to acquire, noisy, small-scale, and highly sensitive to preprocessing pipelines, sites, scanners, and cohort differences. As a result, models trained on a fixed dataset often struggle to generalize across heterogeneous clinical scenarios.

Large language models (LLMs) have recently been applied to knowledge-intensive tasks in specialized domains, including medicine~\cite{liu2026medchain,bleg}, e-commerce risk management~\cite{lu2026sherlocktowardsdynamicknowledgeadaptation}, and finance~\citep{wu2023bloomberggptlargelanguagemodel}. In brain science, LLMs are appealing because they can serve as interfaces to heterogeneous textual knowledge, including neuroscience concepts, clinical criteria, and prior case descriptions. However, existing LLM-based brain network studies mostly follow an \emph{LLM-as-enhancer} paradigm. For example, BrainPrompt uses LLM-generated multi-level prompts to enhance conventional GNNs~\citep{xu2025brainprompt}, while BrainGFM transfers foundation-model ideas to graph pre-training via graph masked autoencoding and contrastive learning~\citep{wei2025braingfm}. These approaches improve graph representation learning, but they do not fully exploit LLMs as interactive reasoning engines that can directly analyze a subject-level brain graph, seek external evidence, and produce structured explanations.
A natural question is whether we can directly use an LLM as a predictor and analyst for brain networks. Our preliminary observations suggest that this is non-trivial. Directly prompting a general-purpose LLM with serialized connectome data often leads to three failure modes. First, there exists a \emph{structure-language gap}: a brain network is a non-Euclidean graph with global topology, community organization, and region-level semantics, whereas an LLM receives a sequential text input. Naively linearizing edges produces long and sparse contexts that are difficult for the model to interpret. Second, there exists a \emph{domain knowledge gap}: general-purpose LLMs are not explicitly grounded in subject-specific neuroimaging evidence and may generate biologically unsupported explanations. Third, we observe a \emph{positive prediction bias}: directly prompted LLMs tend to over-predict disorder-related classes, resulting in high sensitivity but low precision or specificity. Such overconfident behavior is particularly problematic for medical screening scenarios.

Agentic LLM systems provide a promising direction to address these limitations. Retrieval-augmented generation connects parametric LLMs with external non-parametric memory~\citep{lewis2020retrieval}, while tool-using and reasoning-acting frameworks allow LLMs to interact with external environments and APIs~\citep{yao2023react,schick2023toolformer,yao2023tree}. However, directly applying generic agent frameworks to brain network analysis is insufficient. Unlike open-domain question answering, the evidence required here is graph-structured, subject-specific, and biologically constrained. The agent must first transform a connectome into a topology-grounded reasoning state, then retrieve relevant neuroscience knowledge and comparable cases, and finally produce a calibrated decision with explicit analytical evidence.

To this end, we propose \textbf{BrainAgent}, a \emph{brain-specific agentic LLM} framework for knowledge-enhanced brain network analysis. As shown in Figure~\ref{fig:teaser}, BrainAgent reformulates brain connectome classification as an iterative process of \emph{understanding}, \emph{retrieval}, \emph{reasoning}, and \emph{reflection}. Given an rs-fMRI brain graph, BrainAgent first invokes brain-specific analytical tools to extract multi-level topological evidence, including region-level, subgraph-level, and global graph features. It then retrieves relevant neuroscience knowledge and task-specific cases through heterogeneous retrieval modules. Finally, the LLM integrates graph evidence, retrieved knowledge, and case memory to generate both a prediction and a structured rationale, followed by a reflection step to mitigate overconfident or inconsistent decisions. Our contributions are summarized as follows:


\begin{itemize}
    \item We formulate brain network analysis as a knowledge-intensive agentic reasoning problem, where LLMs analyze textualized connectomes with external tools and evidence.
    \item We propose \textbf{BrainAgent}, which integrates topology-aware graph understanding, neuroscience knowledge retrieval, case-level retrieval, and reflection into a unified inference-time framework.
    \item Experiments on multiple rs-fMRI datasets show that BrainAgent improves prediction, reduces positive prediction bias, and produces more interpretable reasoning.
\end{itemize}
\section{BrainAgent}
\label{sec:method}


\subsection{Overall Framework}
\label{sec:overall_framework}
\begin{figure}[t]
  \centering
  \includegraphics[width=1.0\linewidth]{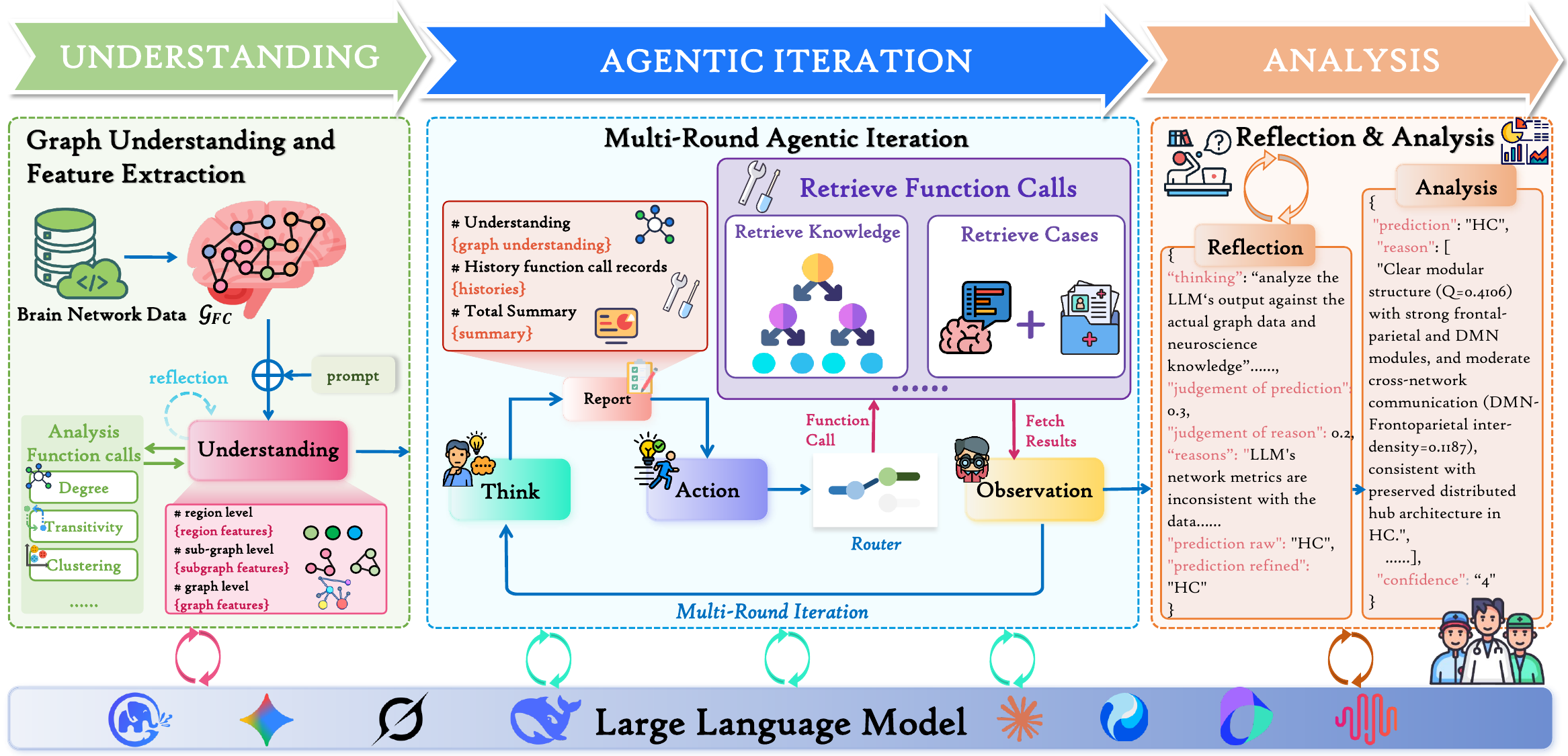}
  \caption{Overall framework of BrainAgent: an understanding module that converts raw brain-network data into compact multi-level structural features, a multi-round agentic iteration module based on a \emph{Think--Report--Action--Observation} loop with analysis tools and external retrieval, and an analysis stage with reflection to verify and refine the final prediction and rationale.}
  \label{fig:brainagent_framework}
\end{figure}

The overall framework of \textbf{BrainAgent} is shown in Figure~\ref{fig:brainagent_framework}. Given an input brain network, BrainAgent first performs an \emph{understanding} stage to extract high-information-density graph descriptors from sparse and verbose raw graph data. This stage converts the original brain network into multi-level structural descriptions, including region-level, subgraph-level, and graph-level features. Compared with directly feeding long edge lists into the LLM, this design reduces token consumption and helps the model better capture brain topology. After understanding, BrainAgent enters a multi-round \emph{agentic iteration} stage. BrainAgent follows a brain-specific \emph{Think--Report--Action--Observation} loop. At each round, the LLM first reasons over the current context, then generates a concise report to summarize key historical information, selects an external tool through a router, and incorporates the returned observation into the next reasoning step. The available tools include brain-network analysis functions, neuroscience knowledge retrieval, and task-specific case retrieval. Finally, BrainAgent produces the final \emph{analysis}, including the predicted label, supporting rationale, and confidence score. To mitigate the overconfidence bias of direct LLM prediction, BrainAgent further uses a reflection module to verify whether the prediction is consistent with graph evidence, retrieved neuroscience knowledge, and similar historical cases.

\subsection{Topology-Aware Graph Understanding}
\label{sec:graph_understanding}

The first step of BrainAgent is to convert the raw brain network into an LLM-readable textual representation. Since the values in a functional brain network naturally indicate connection strengths between brain regions, we represent the brain graph as a sequence of triplets in the form of \verb|(src, dst, val)|, where \verb|src| and \verb|dst| denote two brain regions and \verb|val| denotes their connection strength. To better preserve structural information in the sequential input, we sort the triplets according to the degree of the source brain region, so that structurally important regions appear earlier in the context. However, directly using the serialized graph as input may lead to overly long contexts and unreliable reasoning. To address this, BrainAgent invokes analysis functions to extract compact structural descriptors from the original graph. Specifically, for a brain graph $\mathcal G$, its serialized textual representation is denoted as $T_{\mathcal G}$, and the analysis functions generate corresponding analytical descriptions $T'_{\mathcal G}$:
\begin{equation}
    \mathcal G \rightarrow T_{\mathcal G} 
    \xrightarrow{\mathrm{Analysis\ Functions}}
    T'_{\mathcal G} .
    \label{eq:graph_understanding}
\end{equation}

The analysis functions include two types of measurements. The first type contains general graph properties, such as degree, density, clustering coefficient, and transitivity. The second type contains brain-network-related properties, such as small-worldness, global efficiency, and community organization. The serialized graph and the analytical descriptions are then jointly fed into the LLM:
\begin{equation}
    T_{\mathrm{analysis}}
    =
    F_{\theta}
    \left(
    T_{\mathcal G}, T'_{\mathcal G}; P
    \right),
    \label{eq:analysis_text}
\end{equation}
where $P$ denotes the prompt for graph understanding, and $T_{\mathrm{analysis}}$ denotes the generated topology-aware analysis. 
Here, $F_{\theta}$ is the LLM-based agentic framework defined in the problem formulation, with $\theta$ denoting the parameters of the backbone LLM and its inference-time configuration.
As shown in Figure \ref{fig:brainagent_framework}, the output of this stage is organized into three levels: region-level features, subgraph-level features, and graph-level features. Region-level features describe salient brain regions and hub nodes; subgraph-level features summarize community-level patterns; graph-level features capture whole-brain topological properties. This structured understanding result is used as the initial context for the following agentic reasoning process.

\subsection{Hierarchical Augmented Retrieval for Neural Knowledge}
\label{sec:hark_main}
Accurate agentic reasoning requires not only tool design but also reliable external knowledge. To enhance the domain awareness of the LLM, BrainAgent introduces the \textbf{Hierarchical Augmented Retrieval for Neural Knowledge (HARK)} module. Given the current reasoning context and prompt, HARK retrieves the top-$k$ relevant neuroscience knowledge items and provides them as external evidence for subsequent reasoning. Detailed construction and implementation are provided in Appendix~\ref{app:hark}.

\textbf{Neuroscience knowledge base construction.}
High-quality retrieval depends on a high-quality knowledge base. We construct a neuroscience knowledge base from authoritative manuals and highly cited neuroscience literature. The construction pipeline consists of four steps: raw document collection, knowledge-point extraction, LLM-based filtering, and manual verification. Since BrainAgent is evaluated on multiple brain-network analysis tasks, we organize the knowledge base into task-specific knowledge and task-agnostic knowledge. Specifically, for a task $T_i$, the corresponding task-related knowledge set is denoted as $\mathcal{K}^{T_i}$, while the general neuroscience knowledge set is denoted as $\mathcal{K}^{G}$.

\textbf{Hierarchical retrieval.}
BrainAgent retrieves neuroscience knowledge in a hierarchical manner. Since all brain networks in our experiments are constructed under the same AAL atlas, each task-specific knowledge item can be associated with a set of relevant brain regions. Therefore, HARK first performs coarse-grained retrieval based on anatomical region overlap and then conducts fine-grained retrieval based on semantic similarity in the text embedding space.

\textbf{Coarse-grained retrieval.}
For task-specific knowledge, each knowledge item is represented as a key-value pair, where the key is the associated brain-region set and the value is the textual knowledge content. Given a query $q$ and a knowledge item $k$, we compute their anatomical overlap by IoU:
\begin{equation}
    \mathrm{sim}(q,k)
    =
    \frac{
    |R_q \cap R_k|
    }{
    |R_q \cup R_k|
    },
    \label{eq:hark_iou}
\end{equation}
where $R_q$ and $R_k$ denote the brain-region sets involved in the query and the knowledge item, respectively. Knowledge items whose IoU scores exceed a predefined threshold are retained as candidates.

\textbf{Fine-grained retrieval.}
After coarse retrieval, HARK further ranks the candidate task-specific knowledge items by semantic similarity in the embedding space and returns the top-$k$ most relevant items. For general knowledge $\mathcal{K}^{G}$, which is not explicitly associated with specific brain regions, HARK directly performs semantic retrieval and returns the top-$k$ results. The retrieved knowledge is then appended to the agent context as external neuroscience evidence, helping BrainAgent reduce hallucination and bridge the domain gap between general-purpose LLMs and brain network analysis.

\subsection{Case Augmented Retrieval via Dual-Modality Recall}
\label{sec:card_main}

In addition to general neuroscience knowledge, BrainAgent also requires task-specific evidence from similar brain-network samples. To this end, we introduce \textbf{Case Augmented Retrieval via Dual-Modality Recall} (\textbf{CARD}), which retrieves relevant historical cases from the support set and uses their reports as in-context evidence. The full construction of the case memory, retrieval database, and graph encoder is provided in Appendix~\ref{app:card} and Appendix~\ref{app:gfm}.

\textbf{Case memory construction.} For each task $t$, we construct a support case memory $\mathcal{D}_t$ from the training split. Each case contains three components: the brain graph, the ground-truth label, and an LLM-generated brain-network report. The report summarizes the major graph patterns of each sample, including region-level hubs, subgraph-level organization, and global topological properties. In this way, CARD builds a retrieval database that contains both graph-structured evidence and textual analytical evidence.

\textbf{Dual-modality recall.}
For a given input brain graph $\mathcal G_i$ and a candidate case $\mathcal G_j$ from $\mathcal{D}_t$, CARD measures their similarity from both graph and text modalities. The graph-level similarity $\alpha^{\mathcal G}_{i,j}$ is computed based on graph representations extracted by a pre-trained brain graph encoder, while the text-level similarity $\alpha^{\mathcal T}_{i,j}$ is computed based on the generated reports. The final retrieval score is defined as:
\begin{equation}
    \tilde{\alpha}_{i,j}
    =
    \beta \cdot \alpha^{\mathcal G}_{i,j}
    +
    (1-\beta) \cdot \alpha^{\mathcal T}_{i,j},
    \label{eq:card_similarity}
\end{equation}
where $\beta$ controls the balance between graph-level and text-level matching. In our implementation, $\beta$ is fixed to $0.5$.

\textbf{Case-enhanced reasoning.}
Based on the hybrid score $\tilde{\alpha}_{i,j}$, CARD returns the top-$k$ most similar cases and their corresponding reports. These retrieved cases help BrainAgent compare the current sample with previously observed HC or disorder-related patterns. The dual-modality design provides finer-grained matching, since graph representations capture topological similarity while textual reports preserve interpretable analytical cues. The retrieved cases are then appended to the agent context as case-level evidence for subsequent reasoning and reflection.

\subsection{Reflection and Final Analysis}
\label{sec:reflection_analysis}

After multi-round agentic iteration, BrainAgent generates the final analysis. The output includes three parts: the predicted label, the supporting rationale, and the confidence score. The rationale is required to explicitly refer to graph evidence, retrieved neuroscience knowledge, and similar cases whenever available. To further reduce the overconfidence bias of direct LLM prediction, BrainAgent applies a reflection module after the initial analysis. The reflection module checks whether the predicted label is consistent with the extracted graph features, retrieved knowledge, and case-level evidence. If the initial reasoning contains unsupported claims, inconsistent evidence, or excessive confidence, the LLM is asked to revise the analysis and adjust the final prediction when necessary. The final output follows a structured format:

\begin{brainagentoutput}
{
  "prediction": "HC",
  "reason": [
    "Strong bilateral and homologous coupling indicates preserved interhemispheric symmetry.",
    "Dense mesoscale modules with closed triads suggest coherent local functional integration.",
    "......"
  ],
  "confidence": "3"
}
\end{brainagentoutput}

This structured output enables BrainAgent to provide not only a prediction result but also an interpretable reasoning process for brain network analysis.
\section{Experiments}

\subsection{Datasets}
\label{sec_exp_dataset}
\begin{table}[t]
\centering
\caption{Statistics of brain network analysis datasets}
\resizebox{0.8\textwidth}{!}{
\begin{tabular}{lccccc}
\toprule
\textbf{Dataset} & \textbf{Task} & \textbf{Samples} & \textbf{Brain Regions ($|\mathcal{V}|$)} & \textbf{Classes} & \textbf{Labels} \\
\midrule
\textbf{ABIDE}         & ASD diagnosis         & 618   & 90 & 2 & \{HC, ASD\} \\
\textbf{ADHD}          & ADHD diagnosis        & 938   & 90 & 2 & \{HC, ADHD\} \\
\textbf{HCP}           & Gender classification & 1,039 & 90 & 2 & \{Male, Female\} \\
\textbf{Rest-meta-MDD} & MDD diagnosis         & 2,165 & 90 & 2 & \{HC, MDD\} \\
\bottomrule
\end{tabular}
}
\label{tab:dataset}
\end{table}
We evaluate BrainAgent on four public real-world rs-fMRI brain network datasets, including \textbf{ABIDE} \cite{2014ABIDE}, \textbf{ADHD} \cite{adhd}, \textbf{HCP} \cite{hcp}, and \textbf{Rest-meta-MDD} \cite{mdd}. The statistics of these datasets are summarized in Table~\ref{tab:dataset}. All brain networks are constructed under the AAL atlas with 90 brain regions, and each dataset corresponds to a binary graph-level classification task. Specifically, ABIDE is used for autism spectrum disorder (ASD) diagnosis, ADHD is used for attention deficit hyperactivity disorder diagnosis, HCP is used for gender classification, and Rest-meta-MDD is used for major depressive disorder (MDD) diagnosis. For each dataset, we randomly select 20\% of the samples as the test set and use the remaining samples for retrieval database construction and model-related preparation.

\subsection{Experimental Settings}
We evaluate BrainAgent on both closed-source and open-source LLMs to examine its general applicability. For closed-source models, we use strong proprietary LLMs, including \textbf{DeepSeek v3.2} \cite{dsv32}, \textbf{Qwen3 Max} \cite{qwen3}, and \textbf{Gemini 3.1}. For open-source models, we adopt \textbf{Qwen3.5-35B} \cite{qwen3.5}, \textbf{Qwen3.5-9B} \cite{qwen3.5}, and \textbf{Gemma4-26B} as representative backbones. For each backbone LLM, we compare the vanilla direct-prompting baseline with BrainAgent. In addition, for closed-source LLMs, we include two widely used reasoning baselines: \textbf{Chain-of-Thought (CoT)} \cite{cot}, which encourages step-by-step reasoning through prompting, and \textbf{Reflection} \cite{reflection}, which asks the model to reconsider and revise its initial prediction. For the model generalization study and ablation analysis, we additionally evaluate recent proprietary LLMs including \textbf{GPT 5.3}, \textbf{Grok 4}, \textbf{Hunyuan 3}, and \textbf{Seed 2.0}. For models without public technical papers, we report their official API/model identifiers and access sources in Appendix~\ref{app:experimental_details}.
All methods are evaluated under the same data split and input protocol. For API-based models, we use the official inference interfaces with a low temperature to reduce randomness. For open-source models, inference is performed locally with standard decoding settings. No supervised fine-tuning is applied to any LLM. BrainAgent only augments the inference process through topology-aware understanding, external retrieval, tool use, and reflection. \footnote{All LLM model API calls involved in this paper are up to April 2026.}

\subsection{Evaluation Metrics}
\label{sec:evaluation_metrics}

Following common evaluation protocols for LLM-based prediction tasks, we evaluate all methods using \textbf{Accuracy}, \textbf{Recall}, and \textbf{Precision}. Accuracy measures the overall proportion of correctly classified samples, Recall measures the proportion of positive samples that are correctly identified, and Precision measures the proportion of truly positive samples among all samples predicted as positive. For Recall and Precision under \textit{pass@k}, we first aggregate the $k$ predictions into a final label following the same pass criterion, and then compute class-wise Recall and Precision based on the aggregated predictions. For experiments involving multiple stochastic LLM runs, we report \textit{pass@k}, where a sample is considered correctly predicted if at least one of the $k$ generated predictions is correct. In our experiments, we report results with $k=1$ and $k=3$ when applicable. These metrics allow us to evaluate both the predictive performance and the over-prediction tendency of LLMs in brain network analysis.

\begin{table*}[t]
\centering
\caption{Performance comparison of three closed-source LLMs, including DeepSeek v3.2, Qwen3 Max, and Gemini 3.1, on four public brain network analysis datasets.}
\label{tab:datasets_results}
\scriptsize
\setlength{\tabcolsep}{2.5pt}
\renewcommand{\arraystretch}{1.05}

\begin{tabular}{@{}p{0.49\textwidth}@{\hspace{0.01\textwidth}}p{0.49\textwidth}@{}}

\resizebox{\linewidth}{!}{
\begin{tabular}{llcccccc}
\toprule
\rowcolor{gray!20}
\multicolumn{8}{c}{\textbf{ABIDE}} \\
\midrule
\textbf{LLM} & \textbf{Method}
& \multicolumn{2}{c}{\textbf{Acc.}}
& \multicolumn{2}{c}{\textbf{Recall}}
& \multicolumn{2}{c}{\textbf{Precision}} \\
\cmidrule(lr){3-4}\cmidrule(lr){5-6}\cmidrule(lr){7-8}
& & \textit{@1} & \textit{@3}
& \textit{@1} & \textit{@3}
& \textit{@1} & \textit{@3} \\
\midrule
\multirow{4}{*}{DeepSeek v3.2}
& Direct & 47.15 & 47.15 & 100.00 & 100.00 & 47.15 & 47.15 \\
& + Reflection & 51.20 & 72.78 & 46.56 & 76.17 & 53.34 & 56.52 \\
& + CoT & 48.06 & 55.37 & 82.10 & 90.19 & 53.58 & 53.58 \\
\rowcolor{brainagentgreen}
& \textbf{+ BrainAgent} & \textbf{58.54} & \textbf{86.18} & 63.69 & 82.76 & \textbf{57.45} & \textbf{59.52} \\
\midrule
\multirow{4}{*}{Qwen3 Max}
& Direct & 47.96 & 47.96 & 95.51 & 95.51 & 50.40 & 50.40 \\
& + Reflection & 50.12 & 71.85 & 48.91 & 78.73 & 51.32 & 63.24 \\
& + CoT & 47.96 & 47.96 & 95.51 & 95.51 & 50.40 & 50.40 \\
\rowcolor{brainagentgreen}
& \textbf{+ BrainAgent} & \textbf{56.91} & \textbf{78.86} & 62.06 & 84.48 & \textbf{53.42} & \textbf{64.40} \\
\midrule
\multirow{4}{*}{Gemini 3.1}
& Direct & 59.35 & 59.35 & 100.00 & 100.00 & 53.70 & 53.70 \\
& + Reflection & 63.41 & 69.11 & 96.55 & 100.00 & 56.67 & 60.42 \\
& + CoT & 59.35 & 59.35 & 100.00 & 100.00 & 53.70 & 53.70 \\
\rowcolor{brainagentgreen}
& \textbf{+ BrainAgent} & \textbf{70.25} & \textbf{82.93} & 96.55 & 100.00 & \textbf{62.22} & \textbf{73.42} \\
\bottomrule
\end{tabular}
}
&
\resizebox{\linewidth}{!}{
\begin{tabular}{llcccccc}
\toprule
\rowcolor{gray!20}
\multicolumn{8}{c}{\textbf{ADHD}} \\
\midrule
\textbf{LLM} & \textbf{Method}
& \multicolumn{2}{c}{\textbf{Acc.}}
& \multicolumn{2}{c}{\textbf{Recall}}
& \multicolumn{2}{c}{\textbf{Precision}} \\
\cmidrule(lr){3-4}\cmidrule(lr){5-6}\cmidrule(lr){7-8}
& & \textit{@1} & \textit{@3}
& \textit{@1} & \textit{@3}
& \textit{@1} & \textit{@3} \\
\midrule
\multirow{4}{*}{DeepSeek v3.2}
& Direct & 37.97 & 37.97 & 100.00 & 100.00 & 37.97 & 37.97 \\
& + Reflection & 49.19 & 81.14 & 63.38 & 77.46 & 47.91 & 51.29 \\
& + CoT & 37.97 & 37.97 & 100.00 & 100.00 & 37.97 & 37.97 \\
\rowcolor{brainagentgreen}
& \textbf{+ BrainAgent} & \textbf{61.50} & \textbf{86.18} & 35.21 & 82.76 & \textbf{49.01} & \textbf{53.92} \\
\midrule
\multirow{4}{*}{Qwen3 Max}
& Direct & 37.97 & 37.97 & 100.00 & 100.00 & 37.97 & 37.97 \\
& + Reflection & 48.66 & 71.12 & 77.46 & 85.24 & 61.76 & 64.37 \\
& + CoT & 37.97 & 37.97 & 100.00 & 100.00 & 37.97 & 37.97 \\
\rowcolor{brainagentgreen}
& \textbf{+ BrainAgent} & \textbf{56.91} & \textbf{76.12} & 63.38 & 88.73 & \textbf{61.76} & \textbf{67.90} \\
\midrule
\multirow{4}{*}{Gemini 3.1}
& Direct & 51.87 & 52.41 & 98.59 & 100.00 & 44.03 & 44.37 \\
& + Reflection & 58.82 & 68.45 & 88.73 & 100.00 & 47.73 & 54.62 \\
& + CoT & 53.48 & 56.68 & 91.55 & 100.00 & 44.52 & 46.71 \\
\rowcolor{brainagentgreen}
& \textbf{+ BrainAgent} & \textbf{69.52} & \textbf{83.96} & 90.14 & 100.00 & \textbf{56.14} & \textbf{70.30} \\
\bottomrule
\end{tabular}
}
\\[1.0em]

\resizebox{\linewidth}{!}{
\begin{tabular}{llcccccc}
\toprule
\rowcolor{gray!20}
\multicolumn{8}{c}{\textbf{HCP}} \\
\midrule
\textbf{LLM} & \textbf{Method}
& \multicolumn{2}{c}{\textbf{Acc.}}
& \multicolumn{2}{c}{\textbf{Recall}}
& \multicolumn{2}{c}{\textbf{Precision}} \\
\cmidrule(lr){3-4}\cmidrule(lr){5-6}\cmidrule(lr){7-8}
& & \textit{@1} & \textit{@3}
& \textit{@1} & \textit{@3}
& \textit{@1} & \textit{@3} \\
\midrule
\multirow{4}{*}{DeepSeek v3.2}
& Direct & 45.41 & 45.41 & 100.00 & 100.00 & 45.41 & 45.41 \\
& + Reflection & 46.38 & 75.19 & 56.84 & 82.19 & 45.41 & 51.90 \\
& + CoT & 45.41 & 45.41 & 100.00 & 100.00 & 45.41 & 45.41 \\
\rowcolor{brainagentgreen}
& \textbf{+ BrainAgent} & \textbf{60.84} & \textbf{80.67} & 54.68 & 74.74 & \textbf{50.17} & \textbf{54.21} \\
\midrule
\multirow{4}{*}{Qwen3 Max}
& Direct & 45.41 & 45.41 & 100.00 & 100.00 & 45.41 & 45.41 \\
& + Reflection & 49.19 & 72.84 & 55.12 & 80.90 & 48.91 & 57.76 \\
& + CoT & 45.41 & 45.41 & 100.00 & 100.00 & 45.41 & 45.41 \\
\rowcolor{brainagentgreen}
& \textbf{+ BrainAgent} & \textbf{55.07} & \textbf{80.26} & 64.21 & 87.37 & \textbf{53.50} & \textbf{60.76} \\
\midrule
\multirow{4}{*}{Gemini 3.1}
& Direct & 55.07 & 55.07 & 100.00 & 100.00 & 50.53 & 50.53 \\
& + Reflection & 59.90 & 70.05 & 100.00 & 100.00 & 55.07 & 64.41 \\
& + CoT & 57.49 & 62.32 & 72.32 & 81.25 & 53.37 & 60.51 \\
\rowcolor{brainagentgreen}
& \textbf{+ BrainAgent} & \textbf{63.68} & \textbf{89.60} & 61.54 & 89.25 & \textbf{55.17} & \textbf{88.30} \\
\bottomrule
\end{tabular}
}
&
\resizebox{\linewidth}{!}{
\begin{tabular}{llcccccc}
\toprule
\rowcolor{gray!20}
\multicolumn{8}{c}{\textbf{Rest-meta-MDD}} \\
\midrule
\textbf{LLM} & \textbf{Method}
& \multicolumn{2}{c}{\textbf{Acc.}}
& \multicolumn{2}{c}{\textbf{Recall}}
& \multicolumn{2}{c}{\textbf{Precision}} \\
\cmidrule(lr){3-4}\cmidrule(lr){5-6}\cmidrule(lr){7-8}
& & \textit{@1} & \textit{@3}
& \textit{@1} & \textit{@3}
& \textit{@1} & \textit{@3} \\
\midrule
\multirow{4}{*}{DeepSeek v3.2}
& Direct & 53.58 & 53.58 & 100.00 & 100.00 & 53.58 & 53.58 \\
& + Reflection & 58.01 & 77.46 & 52.38 & 81.93 & 62.50 & 64.66 \\
& + CoT & 53.58 & 53.58 & 100.00 & 100.00 & 53.58 & 53.58 \\
\rowcolor{brainagentgreen}
& \textbf{+ BrainAgent} & \textbf{62.14} & \textbf{84.29} & 60.67 & 83.19 & \textbf{69.98} & \textbf{72.62} \\
\midrule
\multirow{4}{*}{Qwen3 Max}
& Direct & 51.73 & 51.73 & 81.47 & 81.47 & 53.24 & 53.24 \\
& + Reflection & 59.22 & 70.21 & 66.12 & 84.93 & 65.20 & 67.83 \\
& + CoT & 51.73 & 51.73 & 81.47 & 81.47 & 53.24 & 53.24 \\
\rowcolor{brainagentgreen}
& \textbf{+ BrainAgent} & \textbf{68.20} & \textbf{80.21} & 71.25 & 86.93 & \textbf{67.88} & \textbf{69.11} \\
\midrule
\multirow{4}{*}{Gemini 3.1}
& Direct & 55.76 & 56.58 & 100.00 & 100.00 & 54.72 & 55.24 \\
& + Reflection & 59.82 & 67.90 & 97.41 & 100.00 & 57.36 & 62.53 \\
& + CoT & 55.76 & 56.71 & 100.00 & 100.00 & 54.72 & 55.37 \\
\rowcolor{brainagentgreen}
& \textbf{+ BrainAgent} & \textbf{71.36} & \textbf{81.99} & 86.67 & 100.00 & \textbf{67.94} & \textbf{74.84} \\
\bottomrule
\end{tabular}
}

\end{tabular}
\end{table*}

\subsection{Performance on Different LLMs}
\label{sec:llm_performance}

Table~\ref{tab:datasets_results} reports the results of closed-source LLMs on four public datasets. Overall, direct LLM prompting shows limited reliability for brain network analysis. In many cases, vanilla LLMs tend to predict most samples as positive, leading to very high recall but poor accuracy and precision. This indicates a clear positive prediction bias when general-purpose LLMs are directly applied to serialized brain-network data. CoT provides only marginal improvements, while Reflection is more effective by correcting some initial reasoning errors. In contrast, BrainAgent consistently improves accuracy and precision across different datasets and backbone LLMs. The gains are especially evident under \textit{pass@3}, showing that multi-round tool interaction, knowledge retrieval, case retrieval, and reflection help the LLM produce more reliable and evidence-grounded predictions. Nevertheless, the relatively moderate \textit{pass@1} performance also suggests that training-free LLM agents still have limitations in specialized neuroimaging tasks.

\begin{table*}[t]
    \centering
    \caption{Performance comparison of three open-source LLMs, including Qwen3.5-35B, Qwen3.5-9B, and Gemma4-26B, on four public brain network analysis datasets.}
    \label{tab:open_datasets_results}
    \resizebox{\textwidth}{!}{
    \begin{tabular}{lcccccccccccc}
        \toprule
        \textbf{Methods} 
        & \multicolumn{3}{c}{\textbf{ABIDE}}
        & \multicolumn{3}{c}{\textbf{ADHD}}
        & \multicolumn{3}{c}{\textbf{HCP}}
        & \multicolumn{3}{c}{\textbf{Rest-meta-MDD}} \\
        \cmidrule(lr){2-4}
        \cmidrule(lr){5-7}
        \cmidrule(lr){8-10}
        \cmidrule(lr){11-13}
        & \textbf{Acc.} & \textbf{Recall} & \textbf{Precision}
        & \textbf{Acc.} & \textbf{Recall} & \textbf{Precision}
        & \textbf{Acc.} & \textbf{Recall} & \textbf{Precision}
        & \textbf{Acc.} & \textbf{Recall} & \textbf{Precision} \\
        \midrule
        \textbf{Qwen3.5-35B}
        & 47.97 & 91.38 & 47.32
        & 42.25 & 80.28 & 37.75
        & 52.17 & 49.47 & 47.96
        & 53.12 & 92.24 & 53.63 \\
        
        \rowcolor{brainagentgreen}
\textbf{+ BrainAgent}
        & \textbf{59.35} & 81.03 & \textbf{54.65}
        & \textbf{58.29} & 64.79 & \textbf{46.46}
        & \textbf{58.94} & 61.05 & \textbf{54.72}
        & \textbf{61.89} & 90.09 & \textbf{59.54} \\
        
        \midrule
        \textbf{Qwen3.5-9B}
        & 41.46 & 77.59 & 43.27
        & 39.25 & 78.87 & 36.36
        & 49.51 & 49.47 & 45.63
        & 52.66 & 94.40 & 53.28 \\
        
        \rowcolor{brainagentgreen}
\textbf{+ BrainAgent}
        & \textbf{52.85} & 81.03 & \textbf{50.00}
        & \textbf{51.34} & 77.46 & \textbf{42.31}
        & \textbf{57.49} & 34.74 & \textbf{55.93}
        & \textbf{58.43} & 78.02 & \textbf{58.39} \\

                \midrule
        \textbf{Gemma4-26B}
        & 49.59 & 62.07 & 47.37
        & 42.78 & 66.20 & 36.15
        & 49.28 & 64.21 & 46.21
        & 50.12 & 72.84 & 54.28 \\
        
       \rowcolor{brainagentgreen}
        \textbf{+ BrainAgent}
        & \textbf{60.98} & 81.03 & \textbf{55.95}
        & \textbf{56.68} & 76.06 & \textbf{45.76}
        & \textbf{59.22} & 75.53 & \textbf{53.79}
        & \textbf{60.97} & 82.76 & \textbf{59.81} \\
        
        \bottomrule
    \end{tabular}
    }
\end{table*}

Table~\ref{tab:open_datasets_results} further reports the performance of open-source LLMs. In addition to the observations from closed-source models, we find that relatively smaller open-source models can also benefit substantially from BrainAgent. Since BrainAgent is a training-free inference framework, this result suggests that competitive performance can be obtained without relying exclusively on the largest available LLM backbone. Instead, by equipping open-source models with topology-aware understanding, external retrieval, and reflection, BrainAgent can effectively inject domain knowledge during inference. This finding also indicates a promising direction for practical deployment: future work can combine BrainAgent with model compression techniques such as distillation and quantization, enabling smaller local models to support privacy-preserving brain network analysis in real clinical scenarios.

\begin{figure}[t]
  \centering
  \includegraphics[width=1.0\linewidth]{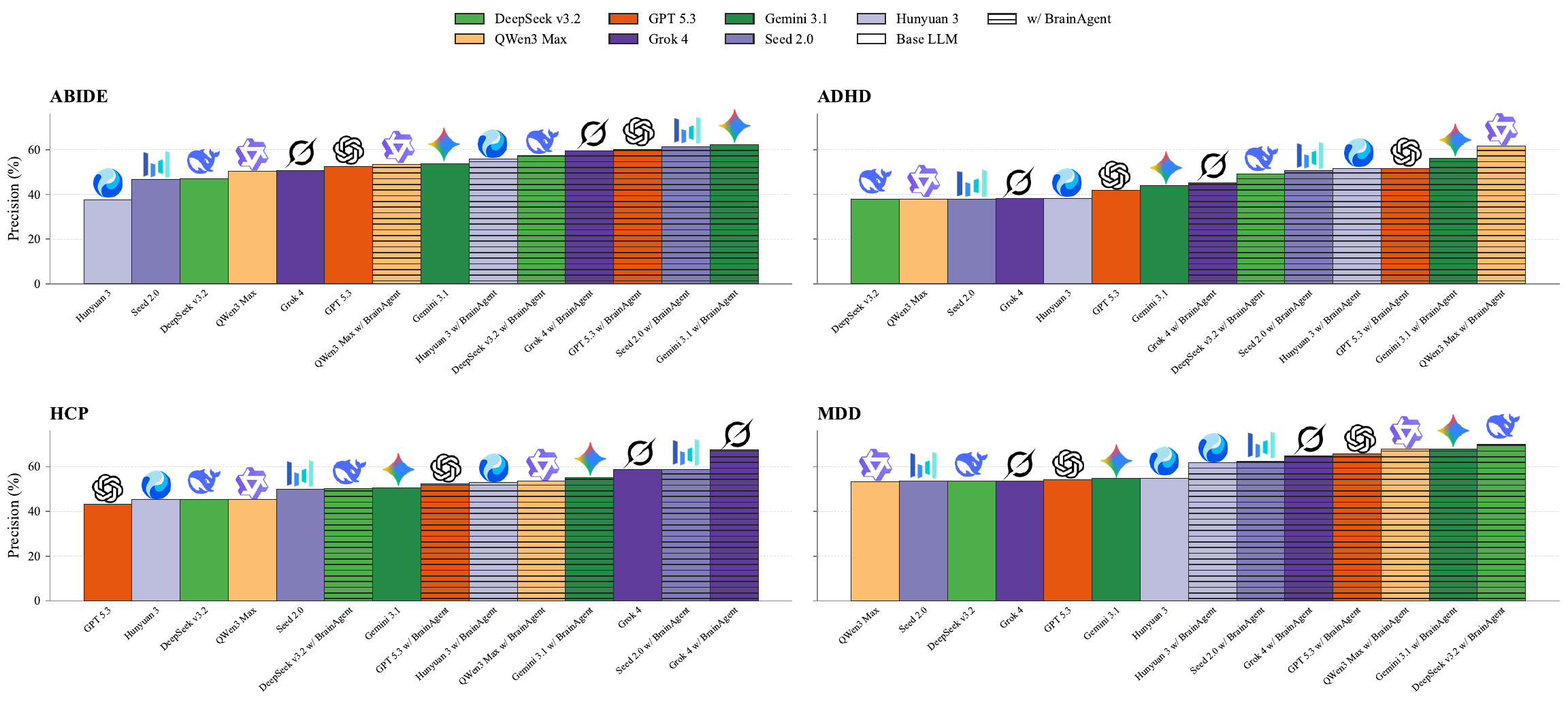}
  \caption{Precision ranking of multiple recent LLMs on four public brain network analysis datasets. BrainAgent consistently improves different backbone LLMs and achieves competitive or leading precision across datasets.}
  \label{fig:llm_ranking}
\end{figure}

\begin{figure}[t]
  \centering
  \includegraphics[width=1.0\linewidth]{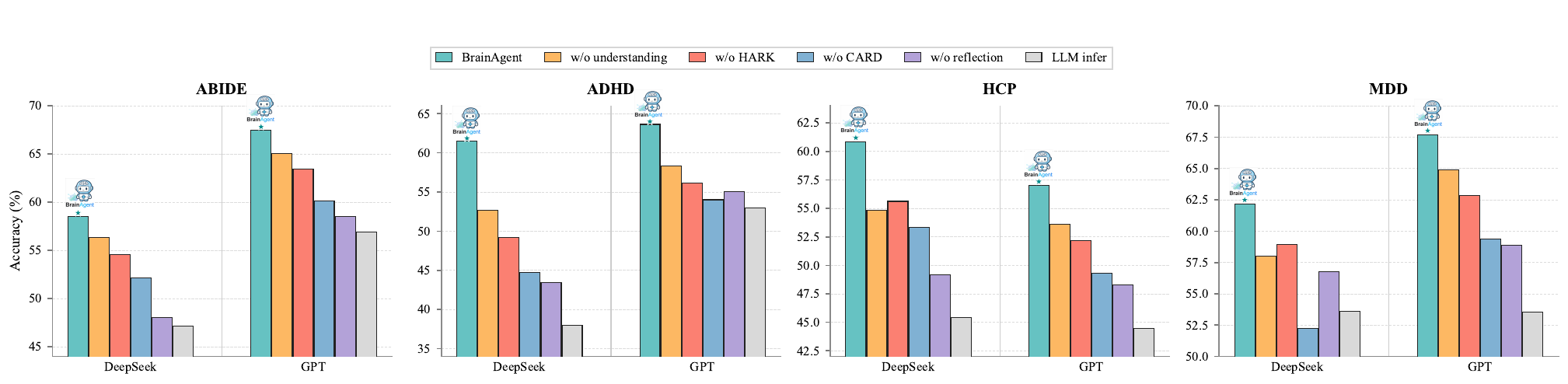}
  \caption{Ablation study of BrainAgent on four public datasets. We report accuracy using DeepSeek v3.2 and GPT 5.3 as backbone LLMs.}
  \label{fig:ablation}
\end{figure}

\subsection{Ablation Studies}
To examine the role of each component, we conduct ablation studies on four public datasets with DeepSeek v3.2 and GPT 5.3 as backbone LLMs. Figure~\ref{fig:ablation} reports the \textit{pass@1} accuracy of BrainAgent and its variants. Removing any module reduces performance, confirming that all components contribute to the final result. The understanding module helps convert long raw graph inputs into compact structural descriptions; HARK provides domain knowledge for reducing the neuroscience knowledge gap; CARD brings task-specific case evidence; and reflection verifies the intermediate reasoning and corrects unsupported conclusions. Among them, removing CARD or reflection usually causes larger drops, highlighting the importance of case-level evidence and self-verification in BrainAgent.

\subsection{Generalization across Recent LLMs}
\label{sec:recent_llms}

To further examine whether BrainAgent is robust to different backbone models, we evaluate it on multiple LLMs and rank their precision on four public datasets. As shown in Figure~\ref{fig:llm_ranking}, BrainAgent consistently improves different base LLMs and often moves them to the top of the ranking. This indicates that the gains of BrainAgent do not come from a specific model, but from the inference-time enhancement introduced by topology-aware understanding, external retrieval, and reflection. Moreover, even when the original base LLM performs moderately, equipping it with BrainAgent can substantially improve its precision, suggesting that BrainAgent effectively reduces unreliable positive predictions and provides a general plug-and-play framework for brain network analysis.

\begin{figure*}[t]
    \centering

    \begin{subfigure}[t]{0.32\textwidth}
        \centering
        \includegraphics[width=\linewidth]{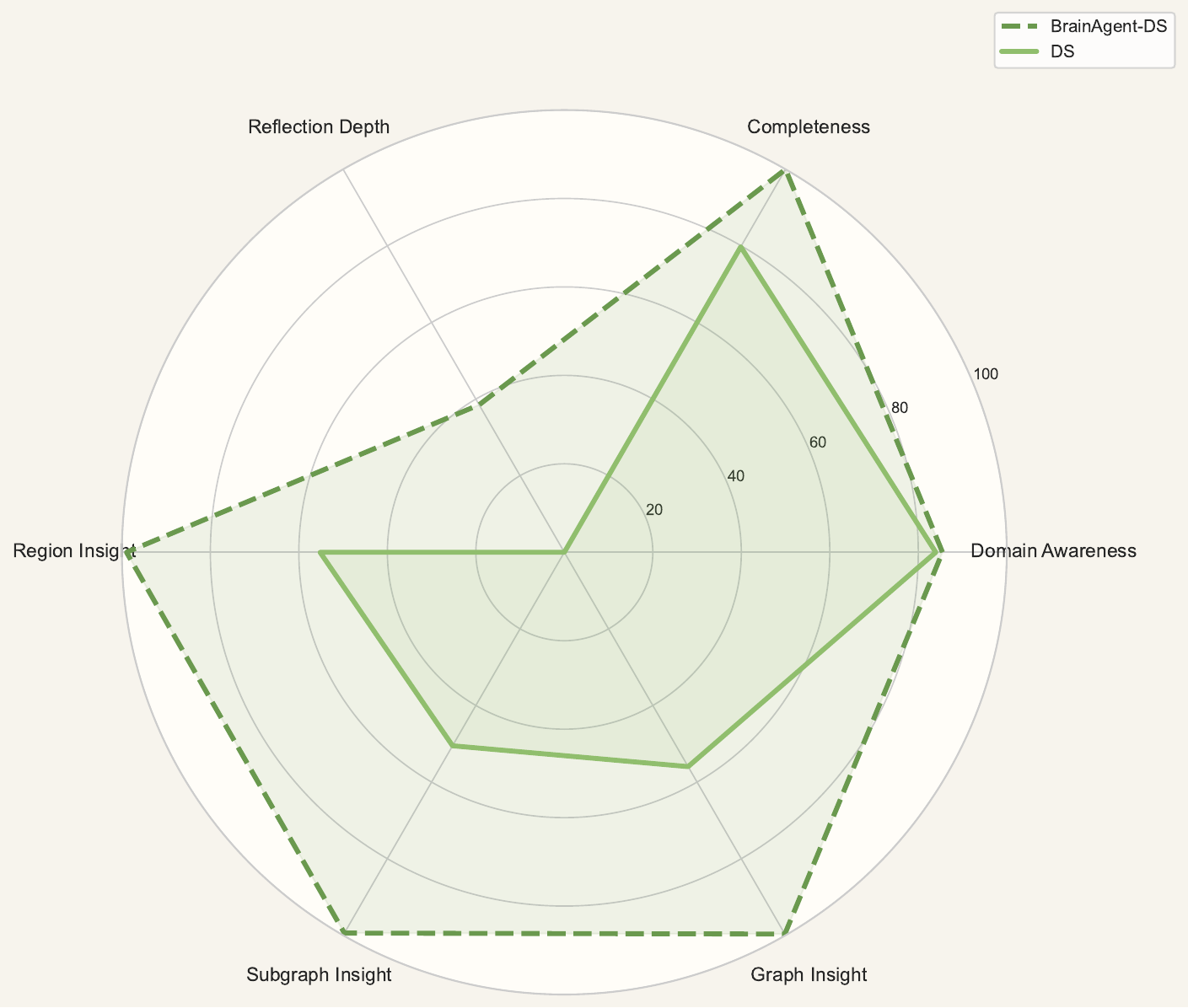}
        \caption{DeepSeek}
        \label{fig:explain_abide_ds}
    \end{subfigure}
    \hfill
    \begin{subfigure}[t]{0.32\textwidth}
        \centering
        \includegraphics[width=\linewidth]{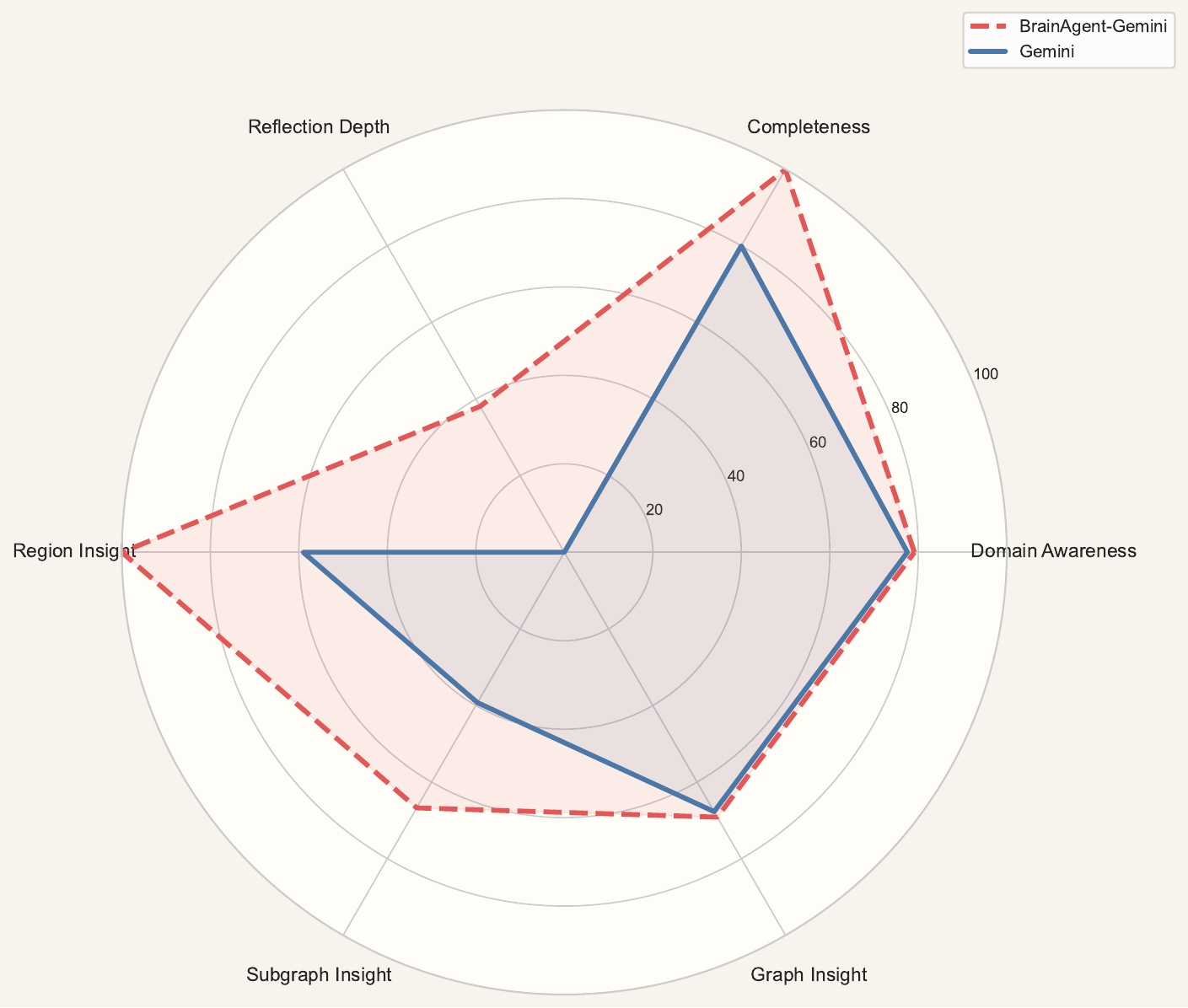}
        \caption{Gemini}
        \label{fig:explain_abide_gpt}
    \end{subfigure}
    \hfill
    \begin{subfigure}[t]{0.32\textwidth}
        \centering
        \includegraphics[width=\linewidth]{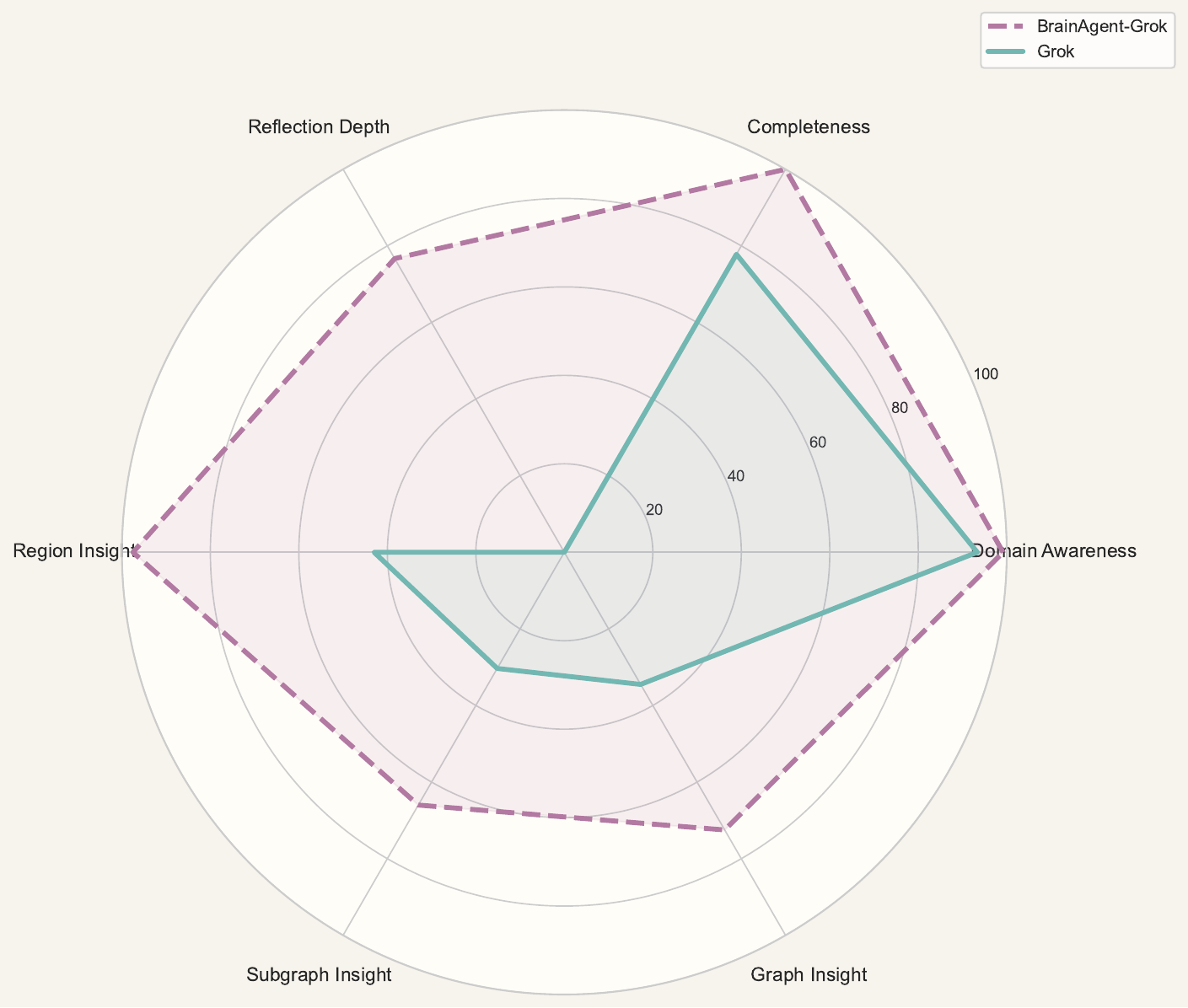}
        \caption{Grok}
        \label{fig:explain_adhd_ds}
    \end{subfigure}
    \hfill
    \begin{subfigure}[t]{0.32\textwidth}
        \centering
        \includegraphics[width=\linewidth]{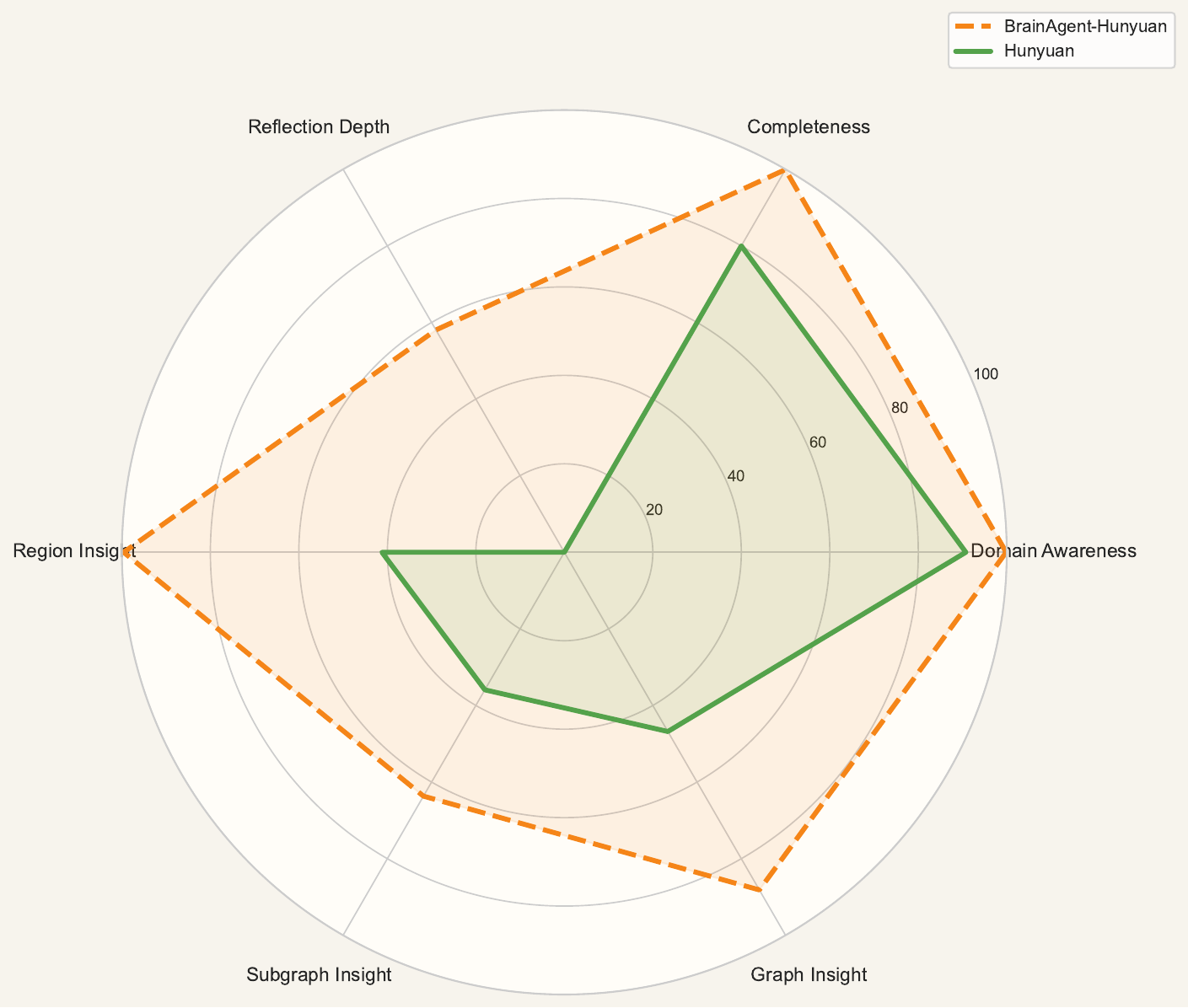}
        \caption{Hunyuan}
        \label{fig:explain_adhd_gpt}
    \end{subfigure}
    \hfill
    \begin{subfigure}[t]{0.32\textwidth}
        \centering
        \includegraphics[width=\linewidth]{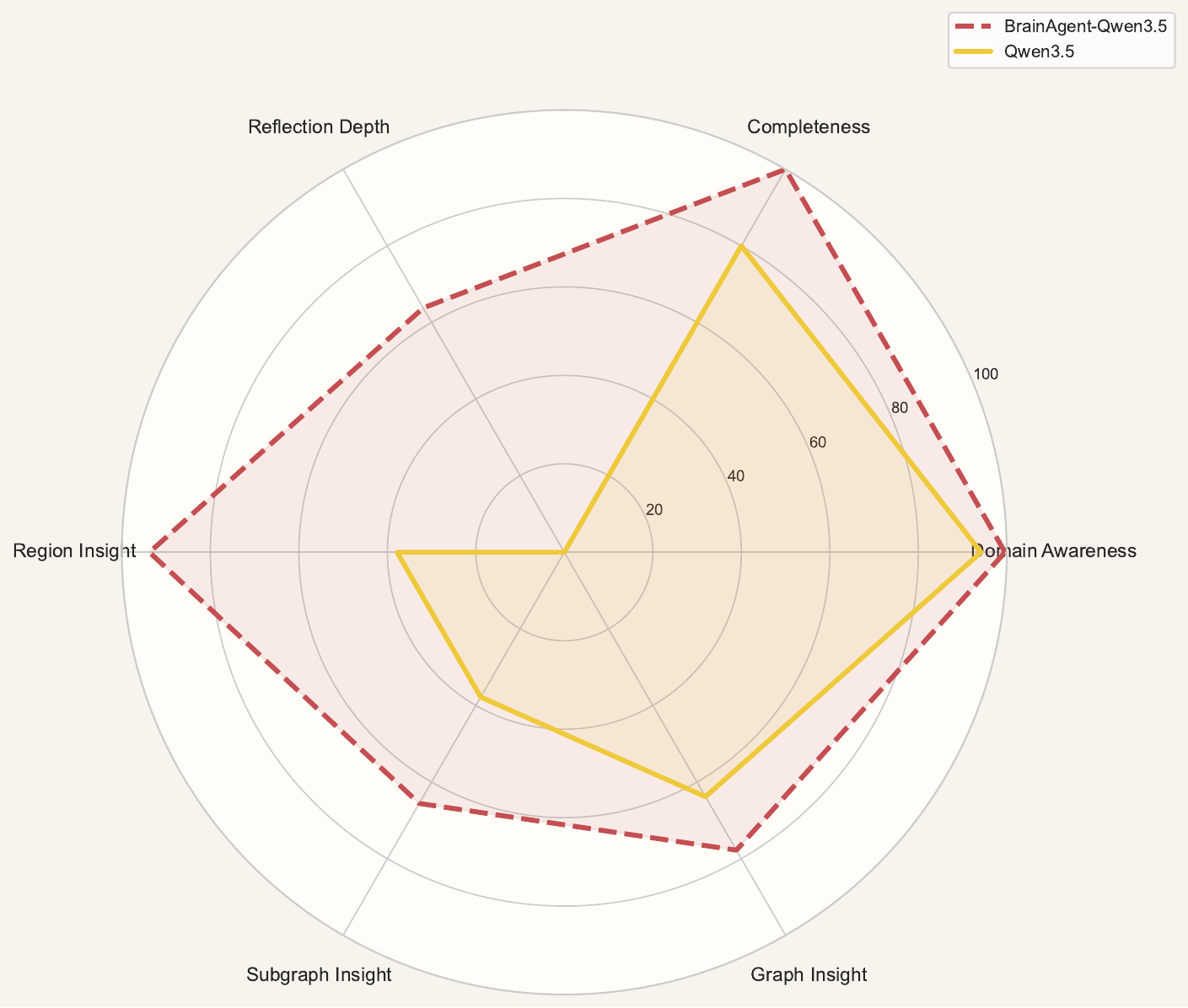}
        \caption{Qwen}
        \label{fig:explain_mdd_ds}
    \end{subfigure}
    \hfill
    \begin{subfigure}[t]{0.32\textwidth}
        \centering
        \includegraphics[width=\linewidth]{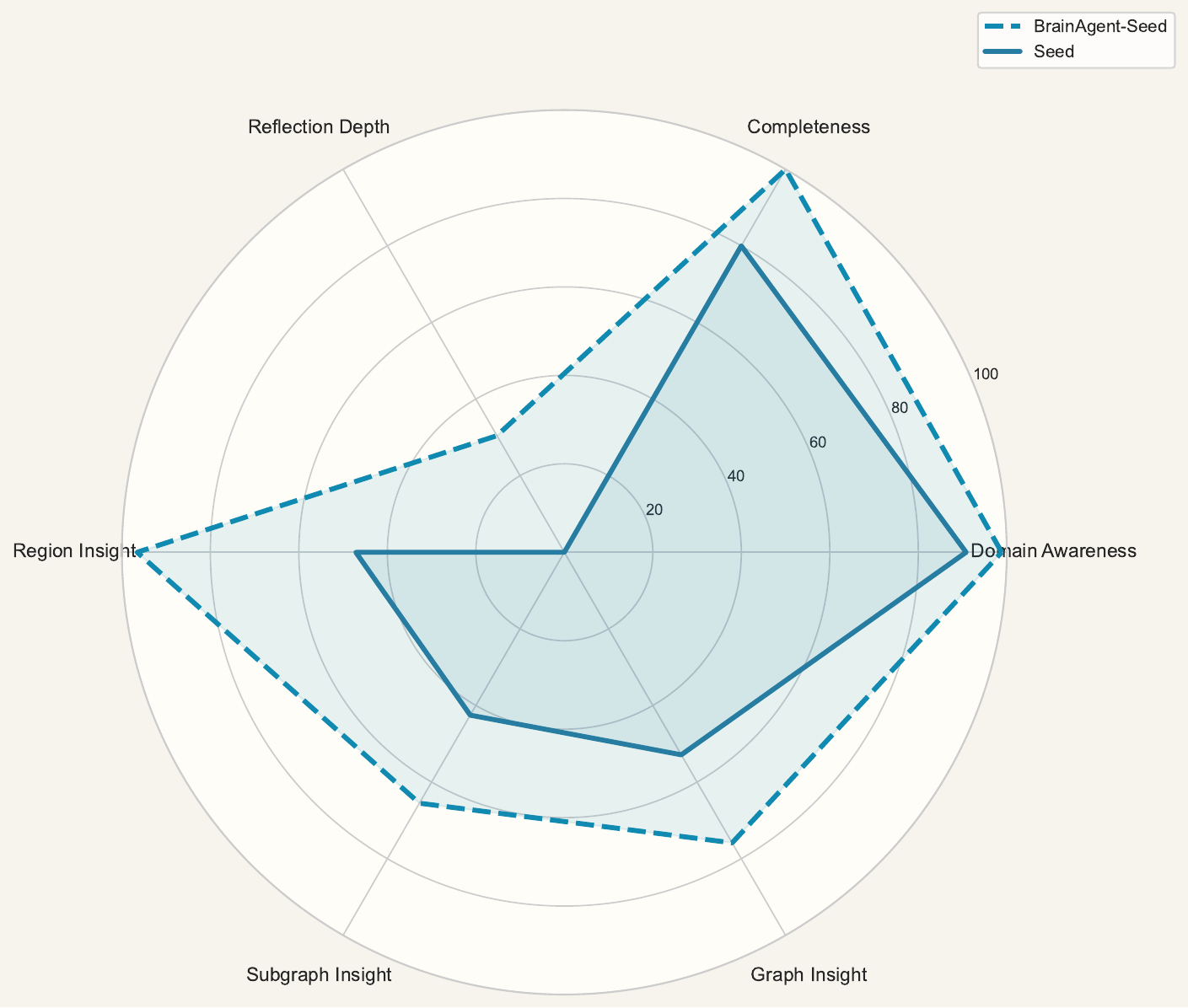}
        \caption{Seed}
        \label{fig:explain_mdd_gpt}
    \end{subfigure}

    \caption{Interpretability comparison between base LLMs and BrainAgent-enhanced LLMs across 6 settings. Each radar plot evaluates explanation quality from 6 dimensions: domain awareness, completeness, reflection depth, region insight, subgraph insight, and graph insight.}
    \label{fig:interpretability_radar}
\end{figure*}

\subsection{Interpretability Analysis}
\label{sec:interpretability}

We evaluate explanation quality from six dimensions: \textbf{Domain Awareness}, \textbf{Completeness}, \textbf{Reflection Depth}, \textbf{Region Insight}, \textbf{Subgraph Insight}, and \textbf{Graph Insight}, as shown in Figure~\ref{fig:interpretability_radar}. These metrics respectively assess whether the explanation is domain-grounded, complete, self-verified, and supported by region-, subgraph-, and graph-level evidence. The overall explainability score is their average. BrainAgent consistently produces more structured explanations than direct inference by decomposing brain networks into multi-level evidence and verifying the rationale through reflection. Detailed reasoning and reflection examples are provided in Appendix~\ref{app:reasoning_examples}.

\section{Conclusion}
\label{sec:conclusion}
We present \textbf{BrainAgent}, a topology-aware agentic LLM framework for interpretable brain network analysis. By combining graph understanding, external knowledge retrieval, case retrieval, and reflection, BrainAgent improves LLM-based connectome prediction across multiple rs-fMRI datasets and backbone models. Experiments further show that BrainAgent mitigates positive prediction bias and produces more structured, evidence-grounded explanations. Future work will explore domain adaptation, better calibration, and extension to multimodal neuroimaging data.

\bibliographystyle{plain}
\bibliography{reference}

\newpage
\appendix
\label{appendix}

\section{Related Works}
\subsection{GNN-based Brain network analysis.}
Graph neural networks (GNNs) have become a widely used backbone for brain network analysis, as they can model non-Euclidean connectome structures through neighborhood aggregation and message passing~\citep{gcn,gat,graphsage}. Existing GNN-based studies for neuroimaging analysis can be roughly grouped into three lines. The first line focuses on designing more expressive graph encoders for functional connectivity networks. For example, connectome-specific architectures and graph neural models are designed to capture higher-order topological patterns and long-range dependencies in brain graphs~\citep{thfcn,brainnpt}. NeuroField-AGL further introduces a neurofield-attentive graph learning mechanism to model functional connectivity for mental disorder diagnosis~\citep{NeuroField}. 

The second line improves interpretability in brain graph learning. BrainGNN learns task-relevant subnetworks and provides interpretable brain-region importance~\citep{li2021braingnn}, while IBGNN and ContrastPool introduce explanation-oriented generators or differentiable pooling mechanisms to identify discriminative substructures~\citep{ibgnn,contrastpool}. These methods improve the transparency of graph classifiers, but their explanations are usually tied to learned graph representations or post-hoc subnetwork selection.

The third line studies multimodal or multi-connectivity brain graph learning. MCRLN integrates structural connectivity, static functional connectivity, and dynamic functional connectivity for MDD diagnosis through structural-functional and static-dynamic fusion~\citep{mcrln}. RH-BrainFS models regional heterogeneity between structural and functional brain networks for multimodal fusion~\citep{rh_brainfs}, M3D-BFS introduces a sample-adaptive dynamic fusion strategy for multimodal brain network analysis~\citep{m3d_bfs}, and NeuroBridge bridges functional and structural brain networks through neural coupling and consistency-guided dynamic graph learning~\citep{neurobridge}. Despite these advances, most existing GNN-based methods remain tied to task-specific training and fixed graph-to-logit prediction. They usually require labeled data for each disease or dataset, provide limited explanations beyond attention weights or post-hoc subnetwork visualization, and are difficult to directly transfer across heterogeneous brain-network tasks. BrainAgent is complementary to these supervised graph learning methods: instead of learning a task-specific classifier, it studies whether an LLM agent can perform training-free, evidence-grounded analysis over subject-level brain graphs.

\subsection{LLM-based methods in neuroscience.}
Pre-trained language models have been widely adapted to biomedical and clinical scenarios. Early studies such as BioBERT~\citep{2020biobert} and BioGPT~\citep{luo2022biogpt} continue pre-training or fine-tuning language models on large-scale biomedical corpora, such as PubMed, to improve domain-specific language understanding. With the development of multimodal LLMs, recent methods further incorporate visual, graph, and structured clinical information. For example, LLaVA-Med builds a vision-language assistant for biomedical image understanding~\citep{li2023llavamed}, while other studies explore medical multimodal fusion with imaging, video, or structured clinical records~\citep{saab2024gemini,zheng2024mmmoe,jiang2024medmoe}.

In brain network analysis, LLMs have recently been used to inject external knowledge or semantic information into graph learning. MMGPL introduces graph prompts for fMRI-related multimodal fusion~\citep{2024mmgpl}. BLEG uses LLMs as fMRI graph enhancers by generating textual augmentation and aligning language-model representations with GNN representations~\citep{bleg}. BrainPrompt further incorporates ROI-level, subject-level, and disease-level prompts to enhance GNN-based neurological condition identification~\citep{xu2025brainprompt}. BrainGFM studies large-scale brain graph pre-training and prompt-tuning across different atlases and disorders~\citep{wei2025braingfm}. These methods show that language models and prompt-based knowledge can benefit brain graph representation learning. However, most of them still follow an \emph{LLM-as-enhancer} paradigm, where the LLM mainly provides prompts, textual features, or auxiliary representations for a trained downstream model. In contrast, BrainAgent treats the LLM as an interactive reasoning agent, allowing it to analyze subject-level brain graphs with topology-aware tools, retrieve neuroscience knowledge and similar cases, and verify its own prediction through reflection.

\section{fMRI Brain Network Construction and Data Preprocessing}
\label{sec_append_preprocess}
\begin{figure*}[!h]
    \centering
    \includegraphics[width=\textwidth]{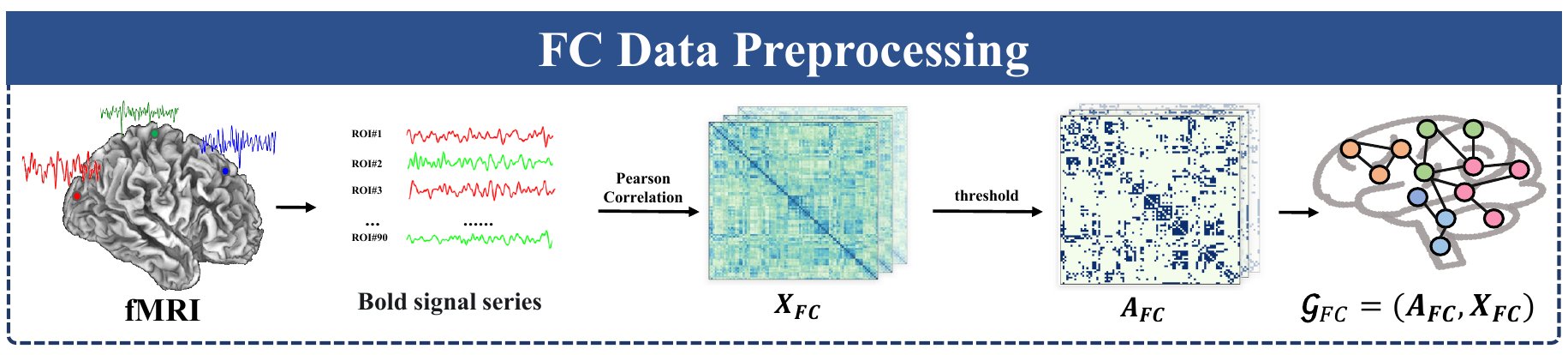}
    \caption{Preprocess of fMRI data and construction for FC dataset.}
    \label{fig_preprocess}
\end{figure*}

The preprocessing pipeline for fMRI data is shown in Figure \ref{fig_preprocess}. The Data Preprocessing Assistant for Resting-State Function (DPARSF) MRI toolkit \cite{DPARSF} is utilized for fMRI preprocessing. Then the average time series are computed for each brain region with AAL template. Pearson correlation is then calculated as functional connectivity matrix, which denotes the feature matrix for FC ($X_{FC}$). Its adjacency matrix ($A_{FC}$) is obtained by thresholding a certain proportional quantization on the functional connectivity matrix.

\section{Details of Input Format and Function Call Settings}

\subsection{Graph-Text Transformation}
The first step in BrainAgent design is to convert graph data into textual representations before feeding them into LLM.
Transforming graph data into formats interpretable by LLMs, as well as enhancing LLMs' capability to comprehend graph topological features, remains a challenging problem \cite{tang2024graphgpt,he2024gretriever}.
On the other hand, compared to graphs in other domains like social networks or molecular chemistry, brain network data exhibits distinct domain-specific properties. According to the preprocess part \ref{sec_append_preprocess}, $\mathcal{E}_{ij}$ represents the connectivity between region $i$ and region $j$ with its value indicating the strength of the connectivity.

Therefore, here we directly represent raw graph data in the form of (\verb|src|, \verb|dst|, \verb|val|) triplets, where \verb|src|, \verb|dst| represent source and target brain regions of each directed edge and \verb|val| stands for connectivity strength. Such transformation ensures complete graph information without information loss.
Moreover, we optimize the ordering of triplets to further enhance LLM's perception of raw graph data \cite{dong2024spreadsheetllm}.
Triplets are sorted in descending order according to connectivity and node degrees, where triplets with large \verb|val| and high degree node are placed in the leading positions.

\subsection{Analysis Function Call Settings}
The basic idea of analysis function calling is to reduce hallucination during inference. For input brain network, LLM tends to conduct certain analysis before final prediction. This process often involves false numerical calculations: They are generated via LLM's next token prediction but not calculated, which often leads to factual errors.
Our goal is to provide these analysis results manually in advance.
Raw graph data, together with corresponding analysis results are fed into BrainAgent for "Understanding", whose output is utilized as input for agent's subsequent reasoning.

The analysis functions in this paper mainly include following two aspects: (a) Basic attributes of the graph, such as edge density, average node degree, average clustering coefficient, transitivity, etc. (b) Statistical attributes related to brain networks, such as small-world property, whole-brain information transmission efficiency, brain community structure attributes, etc.
While calling these analysis function tools, LLM returns functional name and functional description together with analysis results.

\section{Experimental Details}
\label{app:experimental_details}
\subsection{Dataset Details}
\label{app:dataset_details}

We conduct experiments on four public rs-fMRI brain network datasets: ABIDE, ADHD, HCP, and Rest-meta-MDD. All datasets are preprocessed under the AAL atlas, resulting in brain networks with 90 regions of interest (ROIs). Each subject is represented as a functional connectivity graph, where nodes denote brain regions and edges denote functional connectivity estimated from rs-fMRI signals. The detailed statistics are reported in Table~\ref{tab:dataset} in the main paper.

\paragraph{ABIDE.}
The Autism Brain Imaging Data Exchange (ABIDE) dataset \cite{2014ABIDE} is a public multi-site neuroimaging dataset for autism spectrum disorder (ASD) analysis. We use ABIDE for binary classification between healthy controls (HC) and ASD subjects. The dataset contains 618 samples in our experiments.

\paragraph{ADHD.}
The ADHD dataset \cite{adhd} is used for attention deficit hyperactivity disorder diagnosis. It contains rs-fMRI samples from both healthy controls and ADHD subjects. We use 938 samples for binary HC/ADHD classification.

\paragraph{HCP.}
The Human Connectome Project (HCP) dataset \cite{hcp} is used for gender classification. Compared with disease diagnosis datasets, HCP provides a non-disease classification setting and is used to examine whether BrainAgent can generalize beyond clinical diagnosis tasks. The dataset contains 1,039 samples.

\paragraph{Rest-meta-MDD.}
Rest-meta-MDD is a large-scale public resting-state fMRI dataset \cite{mdd} for major depressive disorder (MDD) analysis. It is collected from multiple sites and provides a challenging setting with potential site and cohort heterogeneity. We use 2,165 samples for binary HC/MDD classification.

For each dataset, we randomly split 20\% of the samples as the test set, while the remaining samples are used for retrieval database construction and model-related preparation. No supervised fine-tuning is performed on the LLM backbones.

\subsection{Backbone LLMs and Baselines}
\label{app:llm_baselines}
\begin{table}[t]
\centering
\caption{LLM backbones and access sources used in our experiments.}
\label{tab:llm_versions}
\resizebox{0.8\linewidth}{!}{
\begin{tabular}{lll}
\toprule
\textbf{Model} & \textbf{Provider} & \textbf{Source} \\
\midrule
DeepSeek v3.2 & DeepSeek & Technical report / official API \\
Qwen3 Max & Alibaba Qwen & Technical report / official API \\
Gemini 3.1 & Google & Official API/model interface \\
GPT 5.3 & OpenAI & Official API/model documentation \\
Grok 4 & xAI & Official API/model interface \\
Hunyuan 3 & Tencent Hunyuan & Official API/model interface \\
Seed 2.0 & ByteDance Seed & Official API/model interface \\
Qwen3.5-35B & Alibaba Qwen & Official model release \\
Qwen3.5-9B & Alibaba Qwen & Official model release \\
Gemma4-26B & Google & Official model card/release \\
\bottomrule
\end{tabular}
}
\end{table}
We evaluate BrainAgent with both closed-source and open-source LLMs. For closed-source models, we use DeepSeek v3.2, Qwen3 Max, and Gemini 3.1 as the main backbones. For open-source models, we use Qwen3.5-35B, Qwen3.5-9B, and Gemma4-26B. In addition, we include several recent LLMs in the model generalization study, including GPT 5.3, Grok 4, Hunyuan 3, and Seed 2.0. For models without public technical papers, we report the official API/model identifiers used in our experiments, with access sources summarized in Appendix~\ref{app:experimental_details}.

For each backbone LLM, we compare BrainAgent with direct prompting. For closed-source models, we additionally compare with two common reasoning baselines:
\begin{itemize}
    \item \textbf{CoT}\cite{cot}: Chain-of-Thought prompting, which encourages the model to produce step-by-step reasoning before prediction.
    \item \textbf{Reflection}\cite{reflection}: A reflection-based baseline that asks the model to reconsider and revise its initial prediction after the first inference.
\end{itemize}

All methods use the same data split and input protocol. BrainAgent does not update the parameters of any LLM backbone and only augments inference through graph understanding, external retrieval, tool use, and reflection.

\subsection{Evaluation Metrics}
\label{app:evaluation_metrics}

We evaluate all methods using Accuracy, Recall, Precision, and \textit{pass@k}. Since all datasets in our experiments are binary classification tasks, we regard the disorder-related class as the positive class for ABIDE, ADHD, and Rest-meta-MDD, and use one class as the positive class for HCP following the same evaluation protocol. Let TP, TN, FP, and FN denote true positives, true negatives, false positives, and false negatives, respectively.

\paragraph{Accuracy.}
Accuracy measures the overall proportion of correctly classified samples:
\begin{equation}
    \mathrm{Accuracy}
    =
    \frac{\mathrm{TP}+\mathrm{TN}}
    {\mathrm{TP}+\mathrm{TN}+\mathrm{FP}+\mathrm{FN}} .
\end{equation}
It reflects the general prediction performance of the model over all test samples.

\paragraph{Recall.}
Recall measures the proportion of positive samples that are correctly identified:
\begin{equation}
    \mathrm{Recall}
    =
    \frac{\mathrm{TP}}
    {\mathrm{TP}+\mathrm{FN}} .
\end{equation}
In disease-related brain network analysis, Recall reflects the model's ability to identify disorder-related subjects. However, high Recall alone does not necessarily indicate reliable prediction, since a model that predicts most samples as positive can also obtain high Recall.

\paragraph{Precision.}
Precision measures the proportion of truly positive samples among all samples predicted as positive:
\begin{equation}
    \mathrm{Precision}
    =
    \frac{\mathrm{TP}}
    {\mathrm{TP}+\mathrm{FP}} .
\end{equation}
Precision is particularly important in our setting because direct LLM inference often exhibits positive prediction bias. A low Precision with a high Recall indicates that the model tends to over-predict disorder-related labels and misclassify many healthy controls as positive.

\paragraph{\textit{pass@k}.}
Following common evaluation protocols for LLM-based prediction, we also report \textit{pass@k}. For each test sample, the LLM is queried $k$ times under the same input setting. A sample is counted as correct under \textit{pass@k} if at least one of the $k$ generated predictions matches the ground-truth label:
\begin{equation}
    \mathrm{pass@}k
    =
    \frac{1}{N}
    \sum_{i=1}^{N}
    \mathbb{I}
    \left[
    \exists j \in \{1,\dots,k\},
    \hat{y}_{i}^{(j)} = y_i
    \right],
\end{equation}
where $N$ is the number of test samples, $y_i$ is the ground-truth label of sample $i$, and $\hat{y}_{i}^{(j)}$ denotes the $j$-th generated prediction. In our experiments, we report \textit{pass@1} and \textit{pass@3} when applicable.

For Accuracy under \textit{pass@k}, we directly use the above criterion to determine whether each sample is correctly classified. For Recall and Precision under \textit{pass@k}, we first aggregate the $k$ generated predictions into the final sample-level prediction according to the same pass criterion and then compute Recall and Precision on the aggregated predictions. These metrics jointly evaluate overall prediction quality, sensitivity to positive samples, and the degree of over-prediction in LLM-based brain network analysis.

\subsection{Detailed Analysis of Closed-Source LLM Results}
\label{app:closed_source_result_analysis}

This section provides a more detailed analysis of the closed-source LLM results on four public datasets, including ABIDE, ADHD, HCP, and Rest-meta-MDD. We use DeepSeek v3.2, Qwen3 Max, and Gemini 3.1 as backbone LLMs, and compare direct prompting, CoT, Reflection, and BrainAgent. The main results are reported in Table~\ref{tab:datasets_results}.
Key findings and conclusive take-away messages are shown in Section \ref{sec_discussion}.

\paragraph{Direct LLM inference is still unreliable for brain network prediction.}
Across the four datasets, direct prompting often shows a typical positive prediction bias. In many cases, the model tends to classify most samples into the positive class, which leads to very high Recall but low Precision and low Accuracy. For example, on ADHD and HCP, direct prompting with DeepSeek and Qwen3 Max obtains nearly perfect Recall but poor Accuracy and Precision. This indicates that the LLM is not truly distinguishing disease-related and control samples, but instead tends to over-predict the positive label. Moreover, repeated inference does not always alleviate this issue, as the results of \textit{pass@1} and \textit{pass@3} are often identical for direct prompting. This suggests that simply sampling multiple outputs is insufficient when the model cannot effectively understand the underlying graph topology.

\paragraph{CoT provides limited improvement.}
Chain-of-Thought prompting encourages the LLM to produce intermediate reasoning before prediction, but its improvement is generally limited in this setting. On several datasets, CoT behaves similarly to direct prompting and still suffers from positive prediction bias. This is likely because the main challenge is not only whether the model can generate a reasoning chain, but whether the reasoning chain is grounded in reliable graph evidence and neuroscience knowledge. When the serialized brain graph is long and difficult to interpret, CoT may produce plausible but weakly grounded explanations, which do not necessarily improve the final prediction.

\paragraph{Reflection is more effective than CoT in many cases.}
Compared with CoT, Reflection usually brings larger gains. This shows that verification and revision are useful for LLM-based brain network analysis. Instead of directly trusting the first prediction, the reflection step asks the model to reconsider whether its conclusion is supported by the input evidence. This helps correct part of the initial reasoning errors and reduces unsupported predictions. However, Reflection alone still relies on the original serialized input and the model's internal knowledge, so its improvement is limited when the model lacks sufficient graph-topological understanding or domain-specific evidence.

\paragraph{BrainAgent reduces positive prediction bias and improves prediction reliability.}
BrainAgent consistently improves Accuracy and Precision across datasets and backbone models. Compared with direct prompting, BrainAgent no longer simply predicts most samples as positive. Instead, it uses topology-aware graph understanding, neuroscience knowledge retrieval, case retrieval, and reflection to produce more evidence-grounded predictions. The improvement is especially clear under \textit{pass@3}, indicating that multi-round agentic interaction can make better use of repeated inference. Importantly, the decrease of Recall in some settings should not be interpreted as performance degradation. Since direct prompting often obtains high Recall by over-predicting the positive class, a moderate decrease in Recall together with higher Accuracy and Precision indicates a more balanced decision boundary.

\paragraph{BrainAgent remains a training-free framework with room for improvement.}
Although BrainAgent achieves clear improvements over direct prompting, CoT, and Reflection, its \textit{pass@1} performance is still moderate on some datasets. This reflects the difficulty of directly applying general-purpose LLMs to specialized neuroimaging tasks without task-specific parameter updates. Unlike supervised brain graph learning methods, BrainAgent does not fine-tune the LLM or train a task-specific classifier. Its advantage lies in training-free transfer, flexible tool use, and interpretable reasoning. Future work may further improve this framework through domain-adaptive fine-tuning, better calibration, or reinforcement learning for agentic decision making.

\section{Detailed Interpretability Analysis}
\label{app:interpretability_analysis}

In this section, we provide a more detailed analysis of the interpretability evaluation. As shown in Figure~\ref{fig:interpretability_radar}, we compare the explanations generated by direct LLM inference and BrainAgent-enhanced inference across six backbone LLMs. The radar plots evaluate explanation quality from six dimensions: \textbf{Domain Awareness}, \textbf{Completeness}, \textbf{Reflection Depth}, \textbf{Region Insight}, \textbf{Subgraph Insight}, and \textbf{Graph Insight}. These dimensions are designed to assess whether the generated explanation is not only correct in prediction, but also grounded, structured, and clinically meaningful.

\paragraph{Evaluation dimensions.}
\textbf{Domain Awareness} measures whether the explanation is grounded in the context of brain-network diagnosis, rather than only describing generic graph structures. For example, explanations that mention ASD/HC comparison, clinical relevance, functional connectivity, disease-related biomarkers, or neuroscience literature receive higher scores. 
\textbf{Completeness} evaluates whether the output contains a clear prediction, confidence score, sufficient supporting reasons, and a structured explanation. 
\textbf{Reflection Depth} measures whether the model explicitly checks its own analysis, prediction, and rationale. For BrainAgent, this score mainly comes from the reflection stage, while direct inference usually receives a low score because it does not include an explicit self-verification process. 
\textbf{Region Insight} evaluates whether the explanation discusses node-level or brain-region-level evidence, such as hubs, degree centrality, frontal regions, limbic regions, or parietal regions. 
\textbf{Subgraph Insight} measures whether the explanation analyzes community-level or functional-network-level evidence, such as DMN, FPN, modules, subgraphs, or within-network connectivity. 
\textbf{Graph Insight} evaluates whether the explanation captures global topological properties, such as small-worldness, global efficiency, path length, density, integration, and segregation. The final explainability score is computed as the average of these six dimensions.

Across the six evaluated LLMs, BrainAgent consistently obtains larger radar areas than direct inference. The improvement is especially clear in \textbf{Reflection Depth}, \textbf{Region Insight}, \textbf{Subgraph Insight}, and \textbf{Graph Insight}. This indicates that BrainAgent does not merely provide a label prediction, but decomposes the input brain network into multi-level evidence. In contrast, direct inference often relies on shallow descriptions, such as simply associating dense connectivity or preserved long-range connections with healthy controls. Such explanations may contain some domain-related words, but they usually lack detailed support from region-level hubs, subnetwork organization, and global graph topology.

\paragraph{Effect of multi-level graph understanding.}
The improvement in Region Insight, Subgraph Insight, and Graph Insight mainly comes from the topology-aware understanding stage. BrainAgent first converts the raw connectome into structured descriptions at three levels. At the region level, it identifies salient nodes, hub regions, and high-degree brain areas. At the subgraph level, it summarizes dense modules, community structures, and functional subnetworks. At the graph level, it describes whole-brain topological properties such as small-world organization, global efficiency, and integration--segregation balance. This multi-level decomposition makes the explanation easier to trace and reduces the risk that the LLM produces unsupported diagnostic claims.

\paragraph{Effect of reflection.}
The radar plots also show that BrainAgent substantially improves Reflection Depth. This is because BrainAgent explicitly performs a reflection step after the initial analysis. During reflection, the model checks whether its prediction and rationale are consistent with the extracted graph evidence, retrieved neuroscience knowledge, and similar cases. This mechanism helps the model detect overconfident or weakly supported conclusions. Direct inference lacks this verification stage, and therefore often produces one-step explanations without checking whether the reasoning is faithful to the actual graph structure.

\paragraph{Case study.}
We further analyze a representative ABIDE sample, denoted as ABIDE\_2558, using Grok as the backbone model. The ground-truth label of this sample is ASD. Direct inference predicts the sample as HC, while BrainAgent correctly predicts it as ASD. The difference lies not only in the final prediction, but also in the quality of the reasoning process. Direct inference mainly reasons from a coarse observation that the graph appears dense and that long-range connectivity seems preserved, and therefore concludes that the sample is more similar to HC. This reasoning path is relatively shallow and does not sufficiently examine subnetwork organization or global topology.

By contrast, BrainAgent provides a more layered explanation. In the understanding stage, it first identifies high-connectivity nodes and hub-like regions. It then analyzes dense subgraphs and community-level structures, and finally relates these observations to graph-level properties such as small-worldness and global integration. In the final analysis, BrainAgent organizes high degree centrality, dense subgraphs, and altered global efficiency into an evidence chain supporting the ASD prediction. The reflection stage further checks whether this conclusion is consistent with the graph evidence and retrieved knowledge. As a result, BrainAgent produces a more complete and verifiable reasoning chain.

\begin{brainagentjsoncase}
{
    "prediction": "ASD",
    "reason": [
        "The ABIDE dataset with fMRI modality and AAL template shows high degree centrality in nodes like 0, 16, and 28, which are associated with frontal and limbic regions, indicating hyperconnectivity patterns commonly linked to ASD rather than HC.",
        "Dense sub-graphs in regions related to executive functions (e.g., nodes 10-19) and emotional processing (e.g., nodes 72-81) suggest altered local clustering and community structures, aligning with known ASD biomarkers as confirmed in the previous analysis and retrieved knowledge on sensorimotor hyperconnectivity.",
        "The overall graph exhibits small-world properties with high connectivity and potential changes in global efficiency, such as increased average path length, which are frequently observed in ASD networks compared to HC, as summarized in the report.",
        "The previous report emphasizes reliance on these confirmed patterns for ASD prediction, including increased connectivity and sub-graph densities, without contradictory evidence from retrieve_case, supporting a diagnosis of ASD based on the task of ASD diagnosis."
    ],
    "confidence": "5"
}
\end{brainagentjsoncase}

\vspace{0.6em}

\begin{directjsoncase}
{
    "prediction": "HC",
    "reason": [
        "The ABIDE dataset with fMRI modality and AAL template shows a brain network with a high number of connections (e.g., numerous edges like (0,1,1.000) and (0,10,1.000)), indicating dense functional connectivity, which is often associated with healthy controls in ASD diagnosis tasks.",
        "Statistical analysis of the graph reveals an extensive degree distribution, with nodes such as 0 connected to multiple regions (e.g., 1, 10, 11), suggesting preserved long-range connectivity that is typically more disrupted in ASD subjects.",
        "The graph's overall structure, with no evident isolated nodes and a balanced distribution of edges across regions, aligns with patterns observed in HC groups, as opposed to the hypoconnectivity in social and cognitive networks commonly seen in ASD."
    ],
    "confidence": "3"
}
\end{directjsoncase}

These results suggest that BrainAgent improves interpretability in two complementary ways. First, it transforms the raw brain graph into hierarchical evidence, including region-level, subgraph-level, and graph-level information. Second, it adds a reflection stage to verify whether the final prediction is supported by the extracted evidence. Therefore, BrainAgent's explanations are more structured, more traceable, and closer to the type of evidence-driven reasoning required in brain network analysis.

\section{Details of Hierarchical Augmented Retrieval for Neural Knowledge (HARK)}
\label{app:hark}

\begin{figure*}[htbp]
    \centering
    \includegraphics[width=\textwidth]{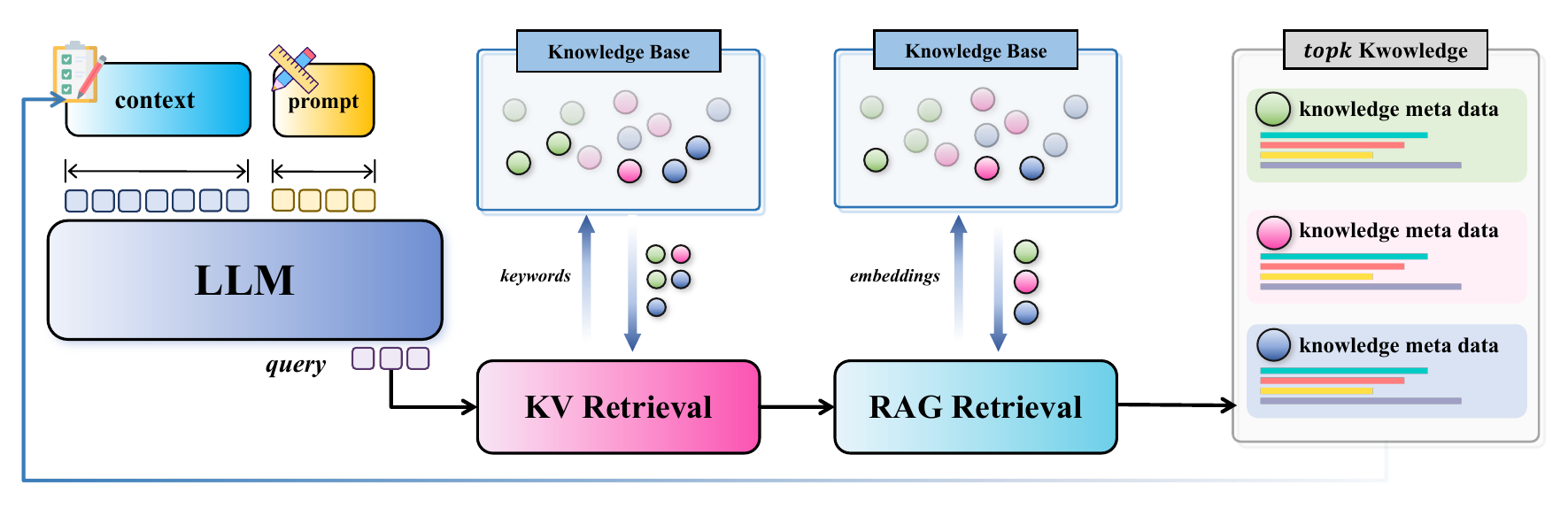}
    \caption{Process of HARK. HARK is a two-stage, hierarchical augmented retrieval for neural knowledge. In the first stage, IoU scores of concerned brain region sets between query and key knowledge is calculated for coarse filtering. Then in the second stage, similarity from embedding level is calculated for fine-grained recall.}
    \label{fig_HARK}
\end{figure*}

We design two distinct function calls aiming to minimize knowledge gap between LLM's general domain and certain task domain. 
HARK injects domain knowledge via general Retrieval Augmented Generation (RAG), narrowing certain domain gap between vanilla LLMs and specialized brain network analysis.

\subsection{Knowledge Base Construction}

\textbf{Knowledge base construction:} HARK is a novel RAG module tailored for brain network analysis. Our knowledge base consists of a set of atomic factual meta knowledge points in neuroscience.
Formally, each meta knowledge data consists of knowledge content and the involved brain regions. We first collect relevant academic papers and professional medical manuals (e.g. DSM5) as source documents according to different tasks (e.g. MDD diagnosis, ADHD diagnosis). Then we prompt LLM to extract useful knowledge points from source documents. The core logic of our prompt follows \cite{mecklenburg2024injecting}'s practice. We employ LLMs to extract and decompose knowledge points from source data, until each decomposed unit only contains core neuroscience knowledge and the corresponding involved brain regions.
Here we conduct document collection and knowledge extraction according to different task types, thus the resulting knowledge set is further categorized by task type. In addition, considering that common knowledge points exist across different brain disease tasks, we also construct a general knowledge set for universal brain network analysis (denoted as ``general'').

\noindent \textbf{Knowledge base refinement:} After the construction of knowledge sets, we adopt different LLM to score, filter, and refine current knowledge data from four dimensions: accuracy, usefulness, linguistic redundancy, and professionalism. For knowledge extraction, we use Deepseek v3.2, and for knowledge refinement, we utilize Kimi k2 with thinking mode.
Furthermore, we map each brain region to its corresponding ID in the AAL template and unify different naming aliases of the same brain region which may cause bias in unprocessed knowledge data.
Finally, the obtained knowledge points are further checked manually.
In this case, we construct a knowledge base consisting of 2000 high-quality knowledge points in total, with average length of each knowledge point 53.2 words. 
Note that in this work the scale of whole knowledge base is limited, and as we focus on complete workflow, improvement of the dataset’s quality and scale can be explored in future works.

\subsection{Hierarchical Augmented Retrieval}
We perform knowledge retrieval on the constructed knowledge base mentioned above.
In summary, as is shown in Figure \ref{fig_HARK} the retrieval framework of HARK integrates semantic-based and representation-based search, realizing a coarse-to-fine two-stage retrieval strategy.
The first stage is semantic-based retrieval, whose core motivation is to filter relevant knowledge points from the corresponding knowledge base for given specific task. We compute the similarity between the brain region set involved in the LLM query ($R_q$) and sets associated with each knowledge point ($R_k$). Then a threshold is adopted for filtering, where one knowledge point will be recalled in this stage if the similarity exceeds the predefined threshold value. We adopt the IoU score as the similarity metric and the threshold is set 0.5. Finally, this stage returns the indices of all recalled knowledge points.

The second stage follows the conventional RAG pipeline. We employ a text representation model to compute the similarity between the query and knowledge points from embedding level, where only the knowledge points recalled in the previous stage are used for similarity calculation. We adopt PubMedBERT \cite{gu2021domain} for text representation, and finally return the top-$k$ knowledge points ranked by similarity. Here we set top-$k$ value as 5. For knowledge from task-specific knowledge base, we follow this two-stage retrieval process. Besides, for knowledge from general knowledge base, we directly retrieve top-$k$ knowledge points from embedding level. In other words, BrainAgent will receive $2 \times \text{top-}k$ knowledge points in total from HARK module.

\section{Details of Graph Foundation Model for Brain Network Data (GFM)}
\label{app:gfm}
In this section, we provide details of the pretraining process for the Graph Foundation Model (GFM).

\begin{figure*}[htbp]
    \centering
    \includegraphics[width=\textwidth]{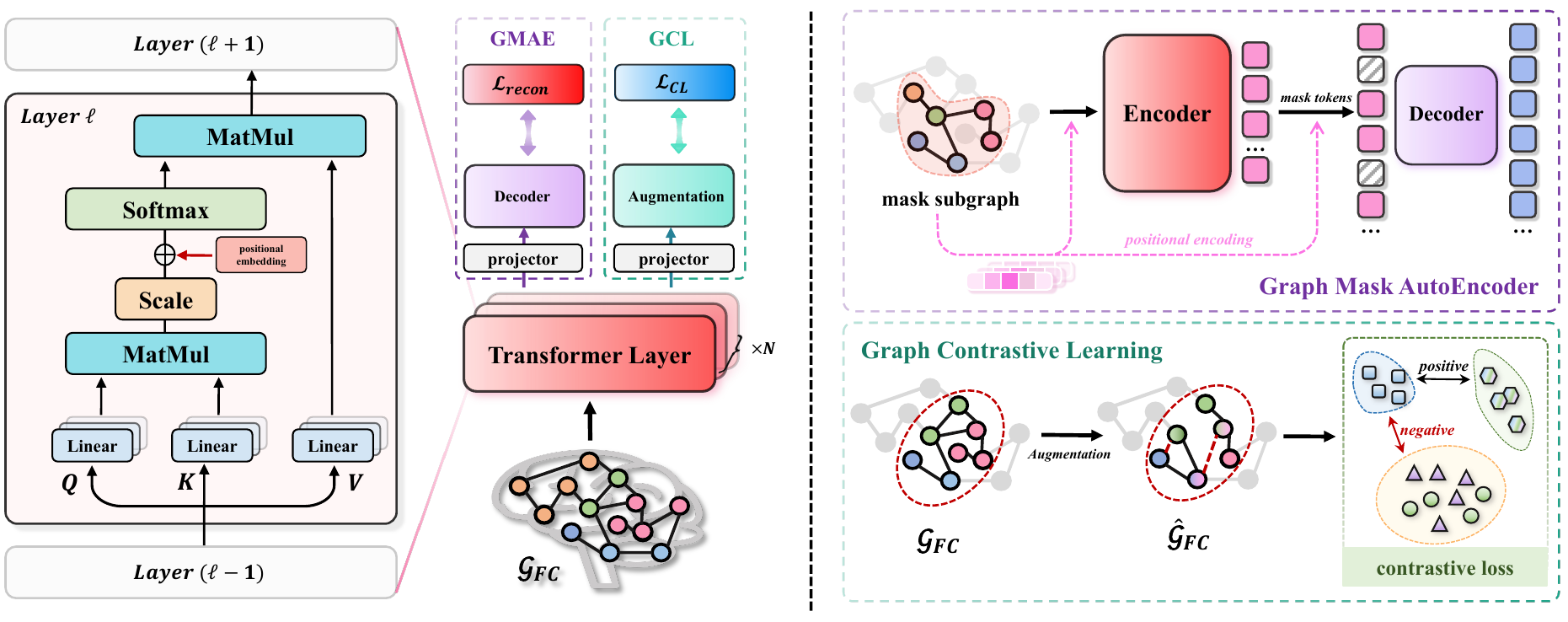}
    \caption{(Left) Structure of Graph Foundation Model. (Right) Training objective of GFM.}
    \label{fig_GFM}
\end{figure*}

\subsection{Overall Pretraining Objective}
The pretraining process consists of two joint optimization objectives: Graph Mask Autoencoder (GMAE) and Graph Contrastive Learning (GCL).

\begin{equation}
    \mathcal{L}_{pretrain} = \mathcal{L}_{recon} + \mathcal{L}_{GCL}
\end{equation}

\subsubsection{Graph Masked Autoencoder (GMAE)}
The main idea of Graph Masked Autoencoder is to reconstruct features of the whole graph from its masked subgraph.
We follow the practice from standard MAE \cite{he2022mae}, by adding random masks on input graph, including node mask and edge mask \cite{hou2022graphmae}.

Formally, for an input graph $\mathcal{G}$ with $N$ nodes, we denote it as $\mathcal{G} = (\textbf{X}, \textbf{A}, \mathcal{V}, \mathcal{E})$. 
For nodes, we randomly select a subset as masked nodes: $\mathcal{V}_M \in \mathcal{V}$. Nodes in $\mathcal{V_M}$ are replaced with learnable vector $\textbf{x}_{[M]}\in \mathbb{R}^{d}$, which is initialized randomly. The node features is then defined as:

\begin{equation}
    \tilde{\textbf{x}}_{i} = 
    \begin{cases}
        \textbf{x}_{[M]}, \quad v_i \in \mathcal{V}_M
        \\
        \textbf{x}_i, \quad \text{otherwise}
    \end{cases}
\end{equation}

For edges, we apply masking by dropping edges randomly. Each edge is dropped independently with probability $p$. Finally we get masked subgraph $\tilde{\mathcal{G}} = (\tilde{\textbf{X}}, \tilde{\textbf{A}})$ with corrupted feature and structure. 
We add positional embeddings (denoted as $\textbf{PE}$) on $\tilde{\textbf{X}}$ which is preprocessed on the whole graph.

The embeddings of un-masked nodes are obtained via encoder (GFM):

\begin{equation}
    \textbf{H}_{encoder} = \text{GFM}\left(\mathcal{\tilde{G}}, \textbf{PE}(\mathcal{G})\right)
\end{equation}

For implementation of GFM, we employ $N$-layer Graph Transformer as our backbone, with each layer composed of multi-head attention and a two-layer feed forward network (FFN).

The input of the decoder is tokens from the whole graph consisting of (a) encoded embeddings for unmasked nodes and (b) masked node tokens. For masked node tokens, they share a uniform learnable vector. Moreover, we add positional embeddings for these masked nodes. For implementation of decoder, we select $M$-layer Transformer for reconstruction, where $N > M$.

The reconstruction target is to predict the raw features of masked nodes, and the loss function is the mean squared error (MSE) between the decoded outputs and raw features:

\begin{equation}
    \mathcal{L}_{recon} = \frac{1}{|\mathcal{V}_M|} \sum_{i =1}^{\mathcal{V}_M} \Big|\Big| \tilde{\textbf{x}}_{i} - \textbf{x}_i \Big|\Big|^2
\end{equation}

\subsubsection{Graph Constrastive Learning (GCL)}
Along with reconstruction task, we adopt contrastive learning method to enhance model's capability under perturbations. Here we directly adopt INFONCE as GCL method.
Formally, we exert perturbations from two different augmented views by randomly dropping out features and edges, and generate augmented graph $\mathcal{G}' = (\textbf{X}', \textbf{A}')$. Then both raw graph $\mathcal{G}_i$ and augmented graph $\mathcal{G}'_i$ are sent into shared GFM encoder and get graph-level embeddings denoted as $z_i, z'_i \in \mathbb{R}^{d}$.
The objective of GCL is to maximize the similarity between embeddings from same graph while keeping the distance from other samples in one Batch $B$:

\begin{equation}
    \mathcal{L}_{GCL} = -\frac{1}{B}\sum_{i=1}^{B}\log \frac{\exp \left( \frac{\text{sim}(z_i, z_i')}{\tau} \right)}{\exp \left( \frac{\text{sim}(z_i, z_i')}{\tau} \right)+\sum_{k \not = i}\exp \left( \frac{\text{sim}(z_i, z_k)}{\tau} \right)}
\end{equation}

\begin{table*}[!ht]
\begin{center}
\begin{tabular}{lc}
\toprule
\textbf{Parameters} & \textbf{Values} \\ 

\midrule
Number of input nodes           &   90 \\
Number of graphs                &   4760 \\ 
Number of GFM layers ($N$)      &   6 \\
Positional encoding (\textbf{PE})   &   \verb|RWSE| \\
\midrule
Hidden dim ($d$)                & 128 \\
Feed forward dim                & 256 \\
Number of attention heads       & 8 \\
Normalization                   &   \verb|LayerNorm()| \\ 
Dropout                         & 0.3 \\
\midrule
Number of GMAE decoder layers ($M$)   & 2 \\
Epochs for pretraining          &  500 \\
Batch size ($B$)                 & 128 \\
Mask ratio for GMAE             & 0.7 \\
Temperature for GCL ($\tau$)    & 1.0 \\

\bottomrule
\end{tabular}
\end{center}
\caption{Parameters for GFM and training process.}
\label{tab_append_gfm_setting}
\end{table*}

\subsection{Details on Experimental Settings}
The backbone of GFM follows the architecture of Graph Transformer, with its encoder constructed by stacking $N$ Transformer blocks (Here, $N=6$). For Graph Transformers, different positional encoding (PE) strategies are often utilized to inject topological information for graphs. Here, we use Random Walk Structural Encoding (RWSE) for implementation, following BrainGFM's practice \cite{wei2025braingfm}.
As we use AAL template for preprocessing, each input brain graph maintains exactly 90 brain regions (nodes). The layer of GFM is 6 and the hidden dim is 128. For each Transformer block, the number of attention heads in each multi-head self-attention layer is 8. More parameter settings are listed in Table \ref{tab_append_gfm_setting}.


\begin{figure*}[!htbp]
    \centering
    \includegraphics[width=0.7\textwidth]{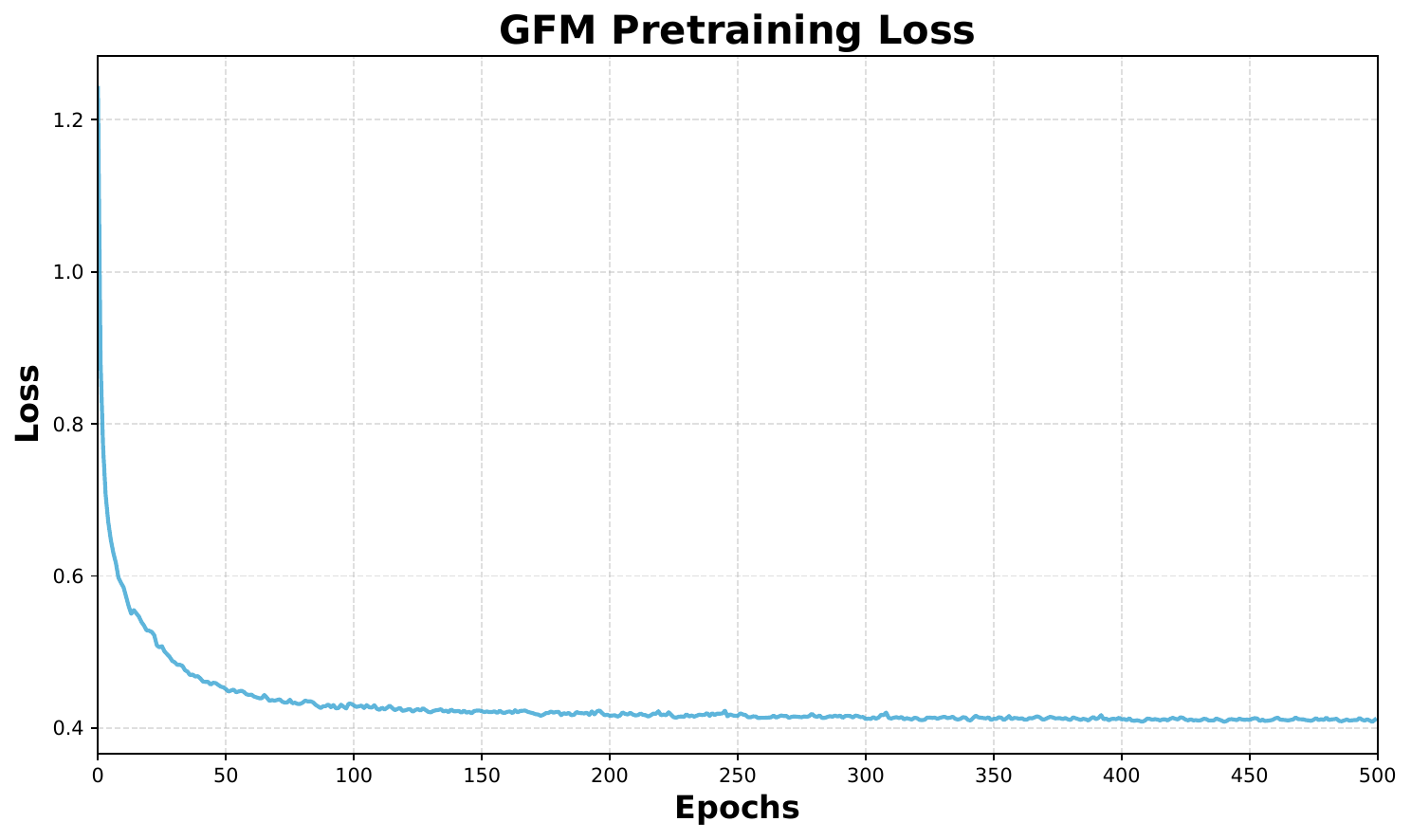}
    \caption{Visualization of loss curve on GFM.}
    \label{fig_gfm_loss}
\end{figure*}

Total epochs of pretraining process is 500, and here the pretraining loss is the sum from GMAE and GCL with no hyper-parameters for adjusting their weights. 
The figure of training loss curve is shown in Figure \ref{fig_gfm_loss}.

\section{Details of Case Augmented Retrieval via Dual-Modality Recall (CARD)}
\label{app:card}

\begin{figure*}[htbp]
    \centering
    \includegraphics[width=\textwidth]{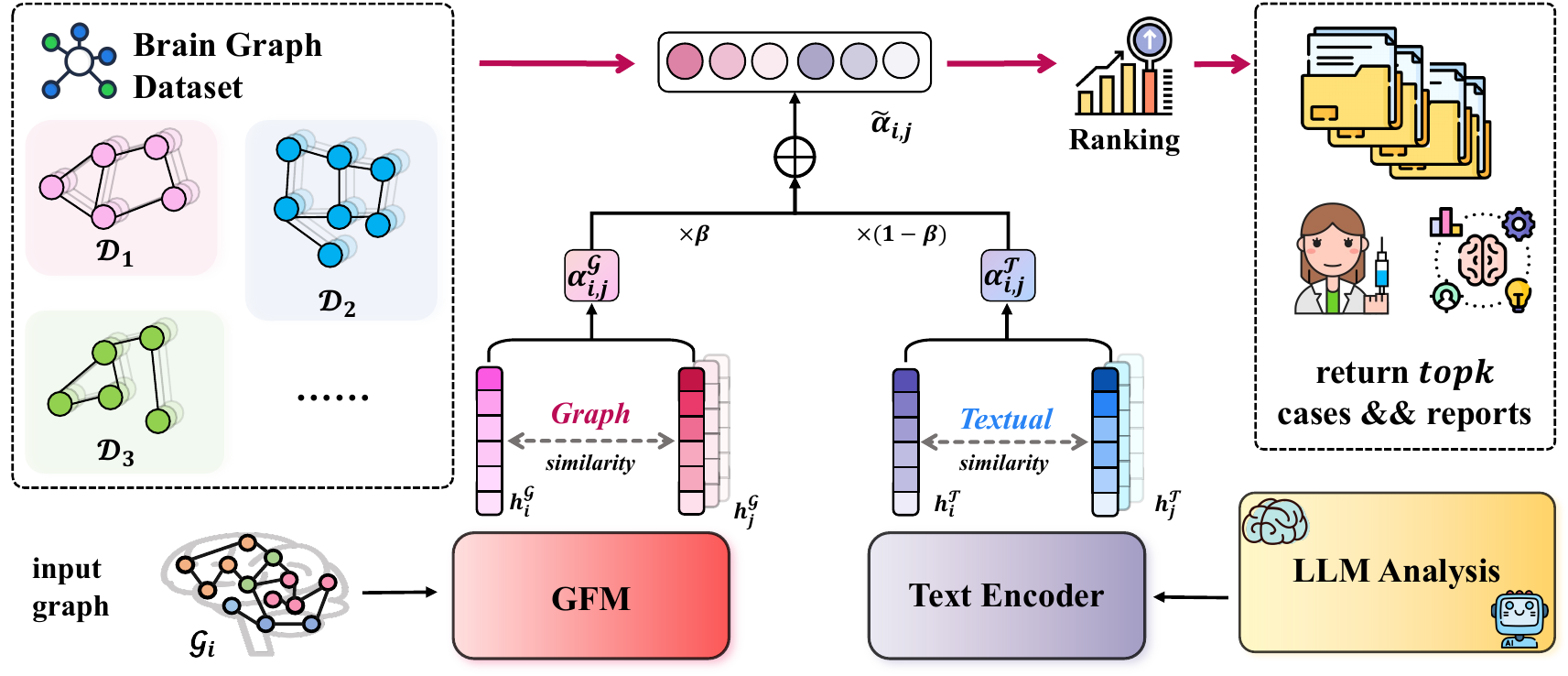}
    \caption{Process of CARD. CARD is a case-level augmented retrieval framwork via dual-modality recall. Graph-level (from GFM) and text-level (from LM) similarities are calculated parallel and the final ranking score is the mixing of two values.}
    \label{fig_CARD}
\end{figure*}

Corresponding to HARK, we design CARD, aiming to achieve knowledge enhancement from relevant samples (cases).
Prior works such as RAGraph~\cite{jiang2024ragraph} and RAG4GFM~\cite{wang2025rag4gfm} combine graph pre-training with retrieval augmentation. Given an input graph, these methods retrieve similar graphs from a constructed memory to improve downstream graph reasoning. CARD follows this general idea but performs retrieval at the subject level for brain network analysis.

Encouraged by this, core innovation of CARD is to compute similarity and retrieve most relevant cases at the sample level. For a given brain network, CARD calculates the similarity between the input and samples in the corresponding dataset, then retrieves the most similar sample data. This further improves the accuracy and interpretability of the subsequent analysis and reasoning of BrainAgent.

As is shown in Figure \ref{fig_CARD}, the overall framework of CARD follows a dual-modality retrieval process. Inspired by RecGPT \cite{yi2025recgpt}, we calculate embedding similarity from a mixing of both graph-level and text-level, leading to finer-grained feature matching.

\subsection{Brain Graph Dataset Construction}
The first step of CARD is to construct a brain graph dataset as case memory based on raw graph format data $\mathcal{G}$. We first split the dataset into training and test sets, following the partitioning setting described in Section \ref{sec_exp_dataset}.
Then for train dataset (also denoted as support dataset), we construct high-quality text-modality report as meta data for case memory.
Similar to HARK's construction pipeline, we first generate report for each data then refine them before final manual check.

\textbf{Dataset construction:} For content generation, we prompt LLM to generate description report for given brain network data. 
Brain networks are organized in a textual form similar to Understanding part, and we also adopt the analytical tools introduced before for augmentation.
The logic of prompt design is roughly consistent with that in Understanding part \ref{sec:graph_understanding}, which requires LLM to analyze brain network graphs and summarize the results into a report. The difference is that we additionally feed the corresponding labels into the LLM. We attribute this to that generating analytical reports based on given labels achieves lower difficulty and higher accuracy than direct analysis \cite{reflection} as it can be regarded as finding proofs instead of pure reasoning. Since this process does not involve test dataset, there is no risk of data leakage. For report generation we utilize Deepseek v3.2 as LLM.
The output format of report strict follows json format which contains information from region level, subgraph level, graph level, case level and a final conclusion.

\textbf{Dataset refinement:} After the LLM generation, we prompt another LLM to check and refine. Refinement process considers the following dimensions: occurrence of factual errors, correctness of the generated analysis report, whether the generated content is redundant and useless, and professionalism of the language style.
Here, we use Kimi k2 with thinking mode as LLM. Moreover, we prompt LLM to output a tag \verb|if_manual| besides from refined content, indicating if the results needs further manual check and correction. Those tagged ``True'' will be checked and modified via human experts in the final stage.

\subsection{Case Augmented Retrieval via Dual-Modality Recall}

Overall architecture of CARD is shown in Figure \ref{fig_CARD}, with the whole pipeline consisting of two parallel modal pathways: graph data and text data. More details are discussed below:

\textbf{Dual-modality embedding: }
The input of CARD contains raw data $\mathcal{G}_i$ and query from BrainAgent $\mathcal{Q}_i$. For graph data, we utilize pretrained GFM to get graph-level embeddings ${h}_i^{\mathcal{G}}$ and for text query we utilize PubmedBERT for textual embeddings ${h}_i^{\mathcal{T}}$, which is shown in Equation \ref{equ_append_card}:

\begin{equation}
\begin{cases}
    \label{equ_append_card}
    h_i^{\mathcal{G}} = \text{Readout}\left( \text{GFM}(\mathcal{G}_i) \right) \in \mathbb{R}^{d_G}
    \\
    h_i^{\mathcal{T}} = \text{PubmedBERT}\left( \mathcal{Q}_i \right) \in \mathbb{R}^{d_T}
\end{cases},
\end{equation}

where $\text{Readout}(\cdot)$ denotes pooling function. Here $d_g=128$ and $d_T=768$.
It is worth noting that the representations of the test dataset are computed in real time ($\mathcal{G}_i$ and $\mathcal{Q}_i$). On the other hand, for training dataset, the corresponding textual data are generated reports. Therefore, we can pre-compute and store these representations locally, thereby further reducing resource consumption during inference phase. For $j$-th data in training dataset, the graph-level embedding is denoted as ${h}_{j}^{\mathcal G}$ and text-level embedding $h_j^{\mathcal{T}}$.

\textbf{Mixing retrieval for graph-text similarity: }
Our retrieval strategy is the mixing of graph and text retrieval process. For given $i$-th data, we calculate cosine similarity for each data $j$ in training dataset. Notice that we partition the data by different tasks at the initial stage. In other words, the bran graph dataset is organized according to task categories, and the retrieval and recall process only needs to be performed within certain dataset. 
Similarity from graph-level and text-level are denoted as $\alpha_{i,j}^{\mathcal{T}}$ and $\alpha_{i,j}^{\mathcal{T}}$ respectively.
The final mixing retrieval score is defined as:
\begin{equation}
    \tilde{\alpha}_{i,j}
    =
    \beta \cdot \alpha^{\mathcal G}_{i,j}
    +
    (1-\beta) \cdot \alpha^{\mathcal T}_{i,j},
\end{equation}
where $\beta$ controls the balance between graph-level and text-level matching. In our implementation, $\beta$ is fixed to $0.5$.
Finally, we retrieve the top-$k$ samples according to the computed similarity $\tilde{\alpha}_{i,j}$ and return their corresponding reports. For CARD, we test results on different top-$k$ ranging in $\{3, 4, 5, 6\}$ and found that best results for top-$k$ is 3, indicating that if the retrieval process is designed useful, only a few reference answers are enough.

\section{Prompt Template}

In this section we provide our designed prompt template for BrainAgent in different parts.

\begin{enumerate}
    \item \textbf{Understanding}: Prompt template for Understanding part is shown in Prompt \ref{prompt_understanding}. The prompt consists of basic data information and statistical results from defined analysis function calls. Output is required as strict json format with region-level, subgraph-level and graph-level analysis.

    \item \textbf{Think and Report}: Prompt template for Think part is shown in Prompt \ref{prompt_think}. BrainAgent is required to perform reasoning before action. Meanwhile, it should also maintain and update a fixed-size report which stores useful history information.

    \item \textbf{Action}: Prompt template for Action part is shown in Prompt \ref{prompt_action}. LLM is prompted to select corresponding function call from available function tools. Meanwhile, concrete queries are required as parameters for different function call. \verb|retrieve_knowledge| stands for HARK module and \verb|retrieve_case| stands for CARD module.

    \item \textbf{Analysis}: Prompt template for Analysis part is shown in Prompt \ref{prompt_analysis}. We prompt LLM to make prediction and give corresponding reasons.

    \item \textbf{reflection}: Prompt template for Reflection part is shown in Prompt \ref{prompt_reflection}. Reflection is made after final analysis for correctness and refinement. LLM is asked to reflect on different generation contents, including LLM analysis, prediction and reasons in  ``Analysis'' part.
    
\end{enumerate}

\clearpage

\begin{promptbox}{Prompt A: Prompt for Understanding}

\colorbox{green!20}{\textbf{\# Role Definition}}\\

You are an expert in neuroscience and machine learning. You can distinguish and analyze difference for given brain network data which takes form of graph.\\

\colorbox{green!20}{\textbf{\# Task Description}}\\

Now give you brain network data which forms as graph, you need to analyze features of the network related to the given data modality.\\

\colorbox{yellow!20}{\textbf{\# Input Data}}\\

\quad - \textbf{data modality}: 

\quad \textcolor{thinkcolor}{\{data\_modality\}}

\quad - \textbf{task}: 

\quad \textcolor{thinkcolor}{\{task\}}

\quad - \textbf{template}: 
AAL template

\quad - \textbf{label}: 
\textcolor{thinkcolor}{\{label\_1\}, \{label\_2\}}

\quad - \textbf{data format description}: 

\quad \textcolor{thinkcolor}{\{data\_description\}}

\quad - \textbf{data}:

\quad \textcolor{thinkcolor}{\{graph\_info\}}

\quad - \textbf{statistical results}:
To further improve accuracy of this part and lower hallunation, some statistical results are provided below. You can directly use them to guide your analysis \textbf{INSTEAD OF CALUCULATING THEM AGAIN ON YOUR OWN.}

\quad \textcolor{thinkcolor}{\{statistical\_results\}}\\

\colorbox{blue!20}{\textbf{\# Requirements}}\\

Your analyze should contain three dimensions: region-level (node), sub-graph-level and graph-level.\\

\textcolor{purple!50}{- \textbf{region-level}}: Analyze the features of each region (node) in the brain network, such as degree, centrality, clustering coefficient, etc. Identify key regions that may play important roles in the network.

\textcolor{purple!50}{- \textbf{sub-graph-level}}: Identify and analyze important sub-graphs or communities within the brain network. Discuss their potential functions and interactions.

\textcolor{purple!50}{- \textbf{graph-level}}: Analyze the overall structure and properties of the brain network.\\

\colorbox{green!20}{\textbf{\# Output Requirements}}\\

Your output must strictly follow json format below:

\vspace{\baselineskip}  

{\color{gray!50}
```json

\{\{

    "region": ["region feature 1", "region feature 2", ...],

    "sub-graph": ["subgraph feature 1", "subgraph feature 2", ...],

    "graph": ["graph feature 1", "graph feature 2", ...]

\}\}

'''
}

\vspace{\baselineskip}  

Now start your analysis.
\label{prompt_understanding}
\end{promptbox}

\begin{promptbox}{Prompt B: Prompt for ``Think'' and ``Report''}
\colorbox{green!20}{\textbf{\# Role Definition}}\\

You are an expert in neuroscience and machine learning. You can distinguish and analyze difference for given brain network data which takes form of graph.\\

\colorbox{green!20}{\textbf{\# Task Description}}\\

Now give you brain network data, you need to give the correct prediction and give your reasons as best as you can. For those knowledge you are not familiar with, you can use tools to support your answer.\\

\colorbox{yellow!20}{\textbf{\# Input Data}}\\
\textcolor{thinkcolor}{\{data\}}\\

\colorbox{yellow!20}{\textbf{\# Last Action and Following Observation}}

\textbf{report: }\\
\textcolor{thinkcolor}{\{report\}}

\textbf{action: }\\
\textcolor{thinkcolor}{\{action\}}

\textbf{observation: }\\ 
\textcolor{thinkcolor}{\{observation\}}\\

\colorbox{blue!20}{\textbf{\# Requirements}}\\

Your main solution should strictly follow 'think' -> 'report' -> 'action' loop. 

\textcolor{blue!50}{- think}: Think is reasoning about the current situation. Your thinking process should combine `input data`, `observation` from latest `action` and `report`.

\textcolor{blue!50}{- report}: After thinking, you should maintain a report. It is the summarization of the current information and should include previous content of report and your current thinking. If the word length is more than 300 words, you should summarize and only extract most valuable key information.

Format of report can follow contents below which MUST BE a **markdown style string**:\\

```

1. Analyze: Your analysis.

2. Action 1: Summarization of action 1 and observation.

...

'''\\

\textcolor{blue!50}{- action}: Your available action must be within these tools below:

\textcolor{observecolor}{\{tools\}}\\

\colorbox{green!20}{\textbf{\# Output Requirements}}\\

Your output must strictly follow json format below. \textbf{DO NOT MODIFY ITS STRUCTURE:}\\
{\color{gray!50}
```json\\
\{\{ \\
	"think": "your thinking process, within 100 words",\\
	"report": "your report, within 300 words",\\
\}\} \\
```\\
}

Now your step is think.

\label{prompt_think}
\end{promptbox}

\begin{promptbox}{Prompt C: Prompt for ``Action''}

\colorbox{green!20}{\textbf{\# Role Definition}}\\

You are an expert in neuroscience and machine learning. You can distinguish and analyze difference for given brain graph network data.\\

\colorbox{green!20}{\textbf{\# Task Description}}\\
Now give you brain network data, you need to give the correct prediction and give your reasons as best as you can. For those knowledge you are not familiar with, use tools to support your answer before giving your final answer.\\

\colorbox{yellow!20}{\textbf{\# Input Data}}\\

\textbf{data:}\\
\textcolor{thinkcolor}{\{data\}}

\textbf{think:}\\
\textcolor{thinkcolor}{\{think\}}

\textbf{report:}\\
\textcolor{thinkcolor}{\{report\}}\\

\colorbox{blue!20}{\textbf{\# Requirements}}\\

Your main solution should strictly follow 'think' -> 'report' -> 'action' loop. 

\textcolor{blue!50}{- think}: Think is reasoning about the current situation. Your thinking process should combine `input data`, `think` content and `report`.

\textcolor{blue!50}{- report}: After thinking, you should maintain a report. It is the summarization of the current information and should include previous content of report and your current thinking. 

\textcolor{blue!50}{- action}: Your available action must be within these types below:\\

\colorbox{blue!10}{\textbf{{\# current tools}}}\\

\verb|<function_call></function_call>| denotes name of the function. NOTE that tool name after it is the name of the tool, DO NOT DIRECTLY OUTPUT function\_call.\\
\verb|<args></args>| denotes arguments of the function. \\
\verb|<description></description>| denotes detailed description and usage of the tool.
-----------------------------\\
\textcolor{blue!50}{\{tools\}}\\

\colorbox{pink!50}{\textbf{{\# Note}}}\\

- For selection of tools, you should combine information from both `data`, `think` and `report` to get more comprehensive information before making decision.\\
- You should select `\verb|analysis_final|` only after `\verb|retrieve_knowledge|` and `\verb|retrieve_case|`.\\

\colorbox{green!20}{\textbf{{\# Output Requirements}}}\\

Your output must strictly follow json format below:

{\color{gray!50}

```json\\
\{\{\\
	"action": "action to take",\\
	"args": \{\{\\
		"arg1": "val1",\\
		"arg2": "val2",\\
		...\\
	\}\}\\
\}\}\\
```\\
}

Now your step is action.
\label{prompt_action}
\end{promptbox}

\begin{promptbox}{Prompt D: Prompt for ``Analysis''}

\colorbox{green!20}{\textbf{\# Role Definition}}\\

You are an expert in neuroscience and machine learning. You can distinguish and analyze difference for given brain network data which takes form of graph.\\

\colorbox{green!20}{\textbf{\# Task Description}}\\

Now give you brain network data which forms as graph, you need to analyze features of the network related to the given data modality before finally making a prediction.\\

\colorbox{yellow!20}{\textbf{\# Input Data}}\\

\textbf{- dataset:}\\
\textcolor{thinkcolor}{\{dataset\}}

\textbf{- data\_modality: } fMRI

\textbf{- template:} AAL

\textbf{- task:} \\
\textcolor{thinkcolor}{\{task\}}

\textbf{- prediction label:}\\
\textcolor{thinkcolor}{\{label\_1\}, \{label\_2\}}

\textbf{- report:} \\
\textcolor{thinkcolor}{\{report\}}\\

\colorbox{blue!20}{\textbf{\# Requirements}}\\

Your analyze should consider complete data and previous analysis, including: 

(1) basic information including `dataset`, `data\_modality`, `template`, `task`, `prediction label`;

(2) `data` which contains statistical analysis of the graph; 

(3) `report` which contains summarization of previous analysis;\\

\colorbox{green!20}{\textbf{\# Output Requirements}}\\

Your output should strictly follow json format below:

{\color{gray!50}

```json\\
\{\{\\
	"prediction": "precition of the data, should be strictly aligned to 'prediction label'",\\
	"reason": ["reason 1", "reason 2", ...],\\
	"confidence": "a number between 1-5 indicating confidence for your judgement. 1 for low confidence, 5 for high and 3 for medium."\\
\}\}\\
```\\
}

Now start your analysis.

\label{prompt_analysis}
\end{promptbox}

\begin{promptbox}{Prompt E: Prompt for "Reflection"}

\colorbox{green!20}{\textbf{\# Role Definition}}\\

You are an expert in neuroscience and machine learning. You can distinguish and analyze difference for given brain network data which takes form of graph. Besides, you are an advanced reasoning agent that can improve based on self refection.\\

\colorbox{green!20}{\textbf{\# Task Description}}\\

Now give you brain network data which forms as graph, as well as its analysis report from LLM. You need to reflect on the generation from LLM based on the input data and other neuroscience knowledge and make refinement and correction if necessary.\\

\colorbox{yellow!20}{\textbf{\# Raw Data info}}\\
\textcolor{thinkcolor}{\{data\_info\}}\\

\colorbox{yellow!20}{\textbf{\# LLM analysis}}\\
\textcolor{thinkcolor}{\{llm\_analysis\}}\\

\colorbox{yellow!20}{\textbf{\# LLM reasoning and prediction}}\\
\textcolor{thinkcolor}{\{llm\_output\}}\\

\colorbox{pink!50}{\textbf{\# Note}}\\

- LLM output include 'llm analysis' 'prediction', 'reason' and 'confidence'.

- Requirements on your reflection:

	- LLM' generation can be reflected in at least these dimension: 

        1. If the analysis from graph data is correct.
		
        2. If the reason is valid based on the data and corresponding analysis. 
		
        3. If the final prediction is correct. 
        
	- Confidence from LLM generation is also a factor indicating llm's own certainty for its prediction.
    
	- To avoid hallucinations, you need to combine your judgement with official neuro-knowledge. You can also refer to external knowledge, tools and resources to improve your reflection.
    
	- After reflection, you should also finally make your own prediction for refinement and correction. 

**REMEMBER that your judgement of the data is based on your own analysis and prediction, as well as refined and reasoning process.**\\

\colorbox{blue!20}{\textbf{\# Output Requirements}}\\

Your output MUST strictly follow json format below:

{\color{gray!50}
```json\\
\{\{ \\
	"thinking": "Key logic of your detailed thinking process",\\
	"judgement of prediction": "Your judgement of the prediction from LLM, a score beween 0 and 1", \\
	"judgement of reason": "Your judgement of the reason from LLM, a score beween 0 and 1", \\
	"reasons": ["Reasons for your reasons on the judgements"], \\
	"prediction\_raw": "The original prediction from LLM", \\
	"prediction\_refined": "Your refined prediction which value can only between [`label1`, `label2`]" \\
\}\} \\
\\```

}

Now start your reflection.

\label{prompt_reflection}    
\end{promptbox}

\section{Visualization of Reasoning Results}
\label{app:reasoning_examples}

To better illustrate how BrainAgent performs interpretable brain-network reasoning, we visualize the intermediate reasoning trajectory of one representative sample from the ABIDE dataset. In this case, we use \textbf{Gemini 3.1} as the backbone LLM. The ground-truth label of this sample is HC, while direct LLM inference predicts it as ASD. In contrast, BrainAgent decomposes the input brain network into multi-level structural evidence, including region-level, subgraph-level, and graph-level descriptions, which provides a more traceable reasoning process for the final prediction.

\begin{reasoningbox}
{
    "region": [
        "High-degree hubs: Node 32 (degree 15), Node 56 (degree 15), Node 57 (degree 16), Node 58 (degree 17), Node 59 (degree 14), Node 61 (degree 14), Node 62 (degree 14), Node 63 (degree 12), Node 84 (degree 14), Node 85 (degree 14) form a densely connected core in posterior brain regions",
        "Frontal lobe nodes (0-23) show moderate connectivity with average degree 5.2, with strongest connections between nodes 10, 12, 13, 18, 19, 60-65",
        "Occipital and parietal nodes (42-55) form a tightly interconnected cluster with average degree 6.8, particularly nodes 42-49 showing near-complete connectivity",
        "Temporal lobe nodes (80-89) show moderate connectivity with average degree 4.3, with nodes 84, 85, 88, 89 serving as connection hubs",
        "Subcortical nodes (70-79) form a distinct module with average degree 5.6, centered around nodes 75, 76, 77",
        "Node 32 (right hippocampus) shows extensive connections to frontal (0,1,10,12,18,19,60-64), temporal (56-69,84,85), and parietal regions",
        "Node 75 (right thalamus) serves as a major subcortical hub with 14 connections to frontal, parietal, and other subcortical regions"
    ],
    "sub-graph": [
        "Frontal network (nodes 0-23): Density 0.42, strongly interconnected with particular strength in prefrontal regions (nodes 10-19)",
        "Default Mode Network components (nodes 31-33, 56-69, 80-81, 84-89): Intra-density 0.395, showing moderate internal connectivity",
        "Occipital-parietal cluster (nodes 42-55): Near-complete subgraph with density 0.92, representing highly synchronized visual processing regions",
        "Subcortical module (nodes 70-79): Density 0.56, centered around thalamic and basal ganglia regions with strong internal connections",
        "Fronto-parietal network connections: DMN-FPN inter-density 0.106, indicating moderate cross-network integration",
        "Temporal lobe cluster (nodes 80-89): Density 0.38, with stronger posterior temporal connections than anterior",
        "Cingulo-opercular network components visible in nodes 24-31 with connections to both frontal and subcortical regions"
    ],
    "graph": [
        "Small-world architecture: ($\sigma$ = 3.0395 indicates strong small-world properties with high clustering and efficient path lengths",
        "High modularity: Q = 0.4696 shows clear community structure with distinct functional modules",
        "Efficient global integration: Global efficiency = 0.5224 suggests good information transfer capability",
        "High local clustering: Average clustering coefficient = 0.5943 indicates strong local connectivity within modules",
        "Characteristic path length = 808988.9551 suggests some long-distance connections but efficient local processing",
        "Network shows typical ASD-related features: preserved small-worldness but altered modular organization compared to HC",
        "DMN integrity: Moderate intra-density (0.395) with reduced long-range connections characteristic of ASD",
        "Fronto-subcortical connectivity: Strong connections between prefrontal regions and thalamus/basal ganglia"
    ]
}
\end{reasoningbox}

This example shows that BrainAgent does not directly rely on a single-step label prediction. Instead, it first converts the raw connectome into structured multi-level evidence. The region-level analysis highlights salient brain regions and hub nodes, the subgraph-level analysis summarizes local community organization, and the graph-level analysis captures whole-brain topological properties. These intermediate outputs make the reasoning process more transparent and provide interpretable evidence for subsequent retrieval, reflection, and final analysis.

\subsection{Visualization of the First-round Reasoning}
\label{app:first_round_reasoning}

After the understanding stage, BrainAgent first summarizes the current structural evidence into a compact report. Based on this report, the agent decides to call the \texttt{retrieve\_knowledge} function to obtain task-related neuroscience knowledge. The retrieved knowledge is returned as the \texttt{observation} field and is used as external evidence for the next reasoning round.

\begin{thinkbox}
{
    "think": "The input data describes a brain network with strong small-world properties ($\sigma$=3.0395), high modularity (Q=0.4696), and efficient global integration. Key features include: a densely connected posterior core (nodes 32,56-59,61-63,84,85), strong occipital-parietal clustering (density 0.92), and preserved but altered DMN connectivity. The report mentions 'typical ASD-related features' including preserved small-worldness but altered modular organization and reduced DMN long-range connections. I need to retrieve knowledge about ASD brain network patterns and compare with similar cases.",
    "report": "## Brain Network Analysis Report\n\n1. **Analyze**: Initial assessment reveals a brain network with:\n   - Strong small-world architecture (($\sigma$=3.0395) indicating efficient information processing\n   - High modularity (Q=0.4696) with distinct functional communities\n   - Densely connected posterior hub regions (nodes 32,56-59,61-63,84,85)\n   - Near-complete occipital-parietal cluster (density 0.92)\n   - Moderate DMN connectivity (intra-density 0.395) with reduced long-range connections\n   - Fronto-subcortical integration via thalamic hub (node 75)\n   - Data suggests 'typical ASD-related features' with preserved small-worldness but altered modular organization"
}
\end{thinkbox}

\vspace{0.8em}

\begin{actionbox}
{
    "action": "retrieve_knowledge",
    "args": {
        "regions_q": ["32", "56", "57", "58", "59", "61", "62", "63", "84", "85", "75"],
        "query": "Brain network shows strong small-world properties (($\sigma$=3.0395), high modularity (Q=0.4696), and efficient global integration. Key features include a densely connected posterior core (high-degree hubs in hippocampus, temporal, and parietal regions), near-complete occipital-parietal cluster (density 0.92), moderate DMN intra-connectivity (0.395) with reduced long-range connections, and strong fronto-subcortical integration via thalamic hub. The report mentions 'typical ASD-related features' with preserved small-worldness but altered modular organization. Need to retrieve knowledge on ASD brain network patterns: how small-worldness, modularity, DMN connectivity, and hub organization differ in ASD compared to healthy controls."
    },
    "observation": [
        "ASD is primarily characterized by repetitive behaviors, with overconnectivity in motor cortex and temporoparietal junction.",
        "ASD is primarily characterized by sensory deficits, with increased connectivity between thalamus and sensory-motor networks.",
        "ASD is primarily characterized by emotional deficits, with underconnectivity between amygdala and inferior parietal lobule.",
        "ASD is primarily characterized by sensory-motor deficits, with underconnectivity between dorsal attention and somatomotor networks.",
        "Default mode network anticorrelated with frontoparietal network in resting-state, consensus from meta-studies.",
        "Inferior frontal network is primarily characterized by language processing, with high connectivity among IFG parts in healthy brains.",
        "Inferior frontal network is primarily characterized by language processing, with high connectivity among IFG parts in healthy brains.",
        "Frontoparietal Network (FPN) is primarily characterized by executive control, with the most robust and reproducible resting-state brain network biomarker being connectivity between middle frontal gyrus and inferior frontal operculum, which is significantly correlated with inhibitory control.",
        "Frontoparietal Network (FPN) is primarily characterized by executive control, with the most robust and reproducible resting-state brain network biomarker being connectivity between middle frontal gyrus and superior parietal lobule, which is significantly correlated with cognitive flexibility."
    ]
}
\end{actionbox}

\subsection{Visualization of the Second-round Reasoning}
\label{app:second_round_reasoning}

The second-round reasoning result is shown below. After receiving neuroscience knowledge from the first round, BrainAgent updates the \texttt{report} field through reflection and further calls the \texttt{retrieve\_case} function. This step retrieves top-$k$ similar samples from the ABIDE case memory (here top-$k$=5), together with their corresponding reports, so that the agent can compare the current brain network with task-specific precedents.

\begin{thinkboxroundtwo}
{
    "think": "The input data shows a brain network with high modularity, small-world properties, and specific hub patterns. The report mentions 'typical ASD-related features' with preserved small-worldness but altered modular organization. The retrieved knowledge includes ASD characteristics like sensory-motor deficits and DMN-FPN anticorrelation. I need to analyze if this network aligns with ASD patterns, focusing on connectivity alterations in DMN, FPN, and subcortical regions.",
    "report": "## Brain Network Analysis Report\n\n1. **Analyze**: Initial assessment reveals a brain network with:\n   - Strong small-world architecture (($\sigma$=3.0395) indicating efficient information processing\n   - High modularity (Q=0.4696) with distinct functional communities\n   - Densely connected posterior hub regions (nodes 32,56-59,61-63,84,85)\n   - Near-complete occipital-parietal cluster (density 0.92)\n   - Moderate DMN connectivity (intra-density 0.395) with reduced long-range connections\n   - Fronto-subcortical integration via thalamic hub (node 75)\n   - Data suggests 'typical ASD-related features' with preserved small-worldness but altered modular organization\n\n2. **Action 1**: Retrieved knowledge on ASD and network biomarkers. Observations indicate ASD involves sensory-motor deficits, DMN-FPN anticorrelation, and executive control linked to FPN connectivity. Key points: ASD shows overconnectivity in motor/temporoparietal areas, underconnectivity in dorsal attention-somatomotor networks, and DMN anticorrelated with FPN."
}
\end{thinkboxroundtwo}

\vspace{0.8em}

\begin{actionboxroundtwo}
{
  "action": "retrieve_case",
  "args": {
    "query_t": "Brain network shows high modularity (Q=0.4696), strong small-worldness (sigma=3.0395), and preserved local clustering (0.5943) but altered modular organization. Key features include densely connected posterior hubs (nodes 32, 56-59, 61-63, 84, 85), a near-complete occipital-parietal cluster (density 0.92), moderate DMN intra-density (0.395) with reduced long-range connections, and strong fronto-subcortical integration via thalamic hub (node 75). These patterns match reported ASD characteristics: preserved small-world architecture but altered community structure, DMN underconnectivity, sensory/visual overconnectivity, and subcortical hub prominence."
  },
  "observation": [
    {
      "similarity_score": 0.857,
      "content": {
        "label": "HC",
        "report": {
          "case_analysis": [
            "For ASD diagnosis using ABIDE fMRI, this network lacks key ASD biomarkers: no fronto-posterior disconnection and no excessive local clustering in sensory regions.",
            "Social brain regions show normal integration, connecting to both DMN and executive networks.",
            "Default Mode Network proxies and Salience Network regions show balanced within- and between-network connectivity, supporting typical anti-correlation dynamics.",
            "The network exhibits flexible modular architecture with well-defined but interconnected communities, unlike ASD's reported overly rigid or fragmented modularity."
          ],
          "conclusion": [
            "The brain network displays neurotypical organization: efficient small-world architecture, balanced integration via hub regions, and well-defined but interconnected functional communities.",
            "Key HC indicators include preserved fronto-temporal integration, symmetric inter-hemispheric connectivity, absence of ASD-typical dysconnectivity patterns, and optimal modularity balance.",
            "Network topology aligns with HC classification from the ABIDE dataset, showing characteristic healthy functional connectivity patterns."
          ]
        }
      }
    },
    {
      "similarity_score": 0.845,
      "content": {
        "label": "HC",
        "report": {
          "case_analysis": [
            "Compared to ASD literature showing DMN disruption, this network shows strong within-DMN-like connectivity in association regions.",
            "The network shows absence of sensory hyper-connectivity patterns: visual and cerebellar regions maintain appropriate segregation without excessive local or long-range connections.",
            "Preserved interhemispheric symmetry in homotopic connections contrasts with reported ASD reductions in corpus callosum connectivity.",
            "Robust hub connectivity in association cortices contradicts ASD findings of weakened integrative hubs and disconnection syndrome patterns."
          ],
          "conclusion": [
            "The brain network displays organizational features consistent with healthy neurotypical connectivity: integrated higher-order networks, segregated sensory processing, and efficient small-world architecture.",
            "Network properties align with HC profiles from ABIDE studies, particularly balanced integration-segregation and preserved interhemispheric coordination.",
            "Absence of connectivity anomalies commonly associated with ASD supports classification as HC."
          ]
        }
      }
    },
    {
      "similarity_score": 0.831,
      "content": {
        "label": "HC",
        "report": {
          "case_analysis": [
            "Compared to ASD networks in ABIDE studies, this graph shows balanced hub distribution, intact long-range connections between frontal and posterior regions, preserved cerebellar integration, and dense cingulate connectivity without fragmentation.",
            "The network shows no evidence of reduced fronto-posterior connectivity, excessive local clustering in sensory regions, or salience network disintegration.",
            "The network topology exhibits efficient global integration and specialized local processing, matching HC profiles."
          ],
          "conclusion": [
            "The functional connectivity network displays characteristics typical of healthy controls: balanced hub architecture, integrated modular organization, preserved long-range connections, and efficient small-world topology.",
            "Specific HC indicators include a high-degree parietal hub, intact frontal-parietal and limbic-temporal integration, cerebellar cortical connectivity, and no evidence of ASD-related dysconnectivity patterns.",
            "While anatomical interpretations require AAL mapping, the topological features strongly align with normative brain network organization observed in HC populations."
          ]
        }
      }
    },
    {
      "similarity_score": 0.828,
      "content": {
        "label": "HC",
        "report": {
          "case_analysis": [
            "The strong posterior DMN cluster connectivity contrasts with ASD findings of DMN hypoconnectivity in ABIDE studies.",
            "Preserved homotopic symmetry aligns with HC patterns rather than ASD-associated interhemispheric dysconnectivity.",
            "Moderate rather than excessive local clustering in sensory regions contradicts ASD hyperconnectivity patterns in primary sensory areas.",
            "Well-distributed hub architecture supports efficient global integration typical of neurotypical networks.",
            "Absence of fragmented community structure or isolated modules is inconsistent with ASD network disintegration phenotypes."
          ],
          "conclusion": [
            "The network exhibits properties consistent with healthy control architecture: modular organization, efficient hubs, and balanced integration-segregation.",
            "Key features contradict ASD biomarkers: robust DMN connectivity, preserved symmetry, and absence of local over-connection patterns.",
            "Graph metrics suggest typical small-world architecture rather than altered topology frequently observed in ASD.",
            "While specific anatomical interpretations require AAL mapping validation, the overall connectivity pattern strongly supports the HC classification.",
            "The binary, thresholded nature of this connectome limits detailed functional interpretation but shows clear differentiation from ASD network phenotypes."
          ]
        }
      }
    },
    {
      "similarity_score": 0.824,
      "content": {
        "label": "HC",
        "report": {
          "case_analysis": [
            "Network shows strong long-range frontal-parietal-temporal connections, inconsistent with ASD hypo-connectivity patterns.",
            "High integration of putative social brain regions suggests preserved social-cognitive network function.",
            "Balanced modularity is observed: distinct functional clusters maintain integration through hub nodes, unlike ASD patterns of reduced between-module connectivity.",
            "Robust interhemispheric connectivity suggested by bilateral hub connections supports typical neural coordination."
          ],
          "conclusion": [
            "The connectivity pattern exhibits hallmark features of healthy brain organization: efficient hub-mediated integration, balanced modularity, and robust long-range connections.",
            "Topological properties, including small-world architecture, rich-club organization, and absence of fragmentation, align with neurotypical resting-state networks.",
            "The network lacks ASD-typical signatures of reduced fronto-posterior connectivity or disrupted default-mode integration, supporting the HC classification.",
            "While anatomical specificity is limited by binary AAL node data, the overall graph topology strongly suggests healthy control connectivity patterns."
          ]
        }
      }
    }
  ]
}
\end{actionboxroundtwo}

\subsection{Visualization of the Third-round Reasoning}
\label{app:third_round_reasoning}

The third-round reasoning result is shown below. After retrieving task-specific cases from the ABIDE case memory, BrainAgent summarizes the conflicting evidence from similar HC and ASD cases. Based on the updated report, the agent performs the final analysis and outputs the predicted label, supporting reasons, and confidence score.

\begin{thinkboxroundthree}
{
    "think": "The retrieved cases strongly indicate HC classification, noting absence of key ASD biomarkers: preserved fronto-posterior integration, robust DMN connectivity, balanced modularity, and no sensory hyperconnectivity. My initial analysis flagged 'typical ASD-related features' from input, but cases contradict this, suggesting the input's note may be misleading. Need to reconcile: input shows strong small-worldness, high modularity, and hub connectivityconsistent with HC per cases. Likely the input's 'ASD-related' comment is a red herring; network topology aligns with HC.",
    "report": "## Brain Network Analysis Report\n\n1. **Analyze**: Initial assessment reveals a brain network with strong small-world architecture , high modularity (Q=0.4696), densely connected posterior hubs, near-complete occipital-parietal cluster (density 0.92), moderate DMN connectivity, and fronto-subcortical integration. Input data mentions 'typical ASD-related features' but this appears inconsistent with detailed topology.\n2. **Action 1**: Retrieved 5 similar cases all labeled HC (similarity scores 0.857-0.824). Key observations: absence of ASD biomarkers-preserved fronto-posterior integration, robust DMN connectivity, balanced modularity, no sensory hyperconnectivity, and intact hub architecture. Cases indicate network exhibits neurotypical organization."
}
\end{thinkboxroundthree}

\vspace{0.8em}

\begin{analysisboxroundthree}
{
    "prediction": "HC",
    "reason": [
        "The network exhibits strong small-world architecture, which are hallmarks of efficient neurotypical brain organization, not the disrupted integration often seen in ASD.",
        "The posterior hub network (nodes 32, 56-59, 61-63, 84, 85) forms a densely connected core with high-degree nodes, indicating preserved hub integrity and efficient long-range communication typical of healthy controls.",
        "The near-complete occipital-parietal cluster (density 0.92) represents normal, highly synchronized sensory processing, contrasting with the sensory hyperconnectivity or hypocommunication often reported in ASD networks.",
        "Moderate Default Mode Network (DMN) intra-density (0.395) and preserved fronto-parietal integration (inter-density 0.106) align with neurotypical connectivity patterns, whereas ASD typically shows reduced long-range DMN connectivity and altered network segregation.",
        "Retrieved similar cases from the report all labeled HC (similarity scores 0.857-0.824) provide strong empirical evidence that this network topology is characteristic of healthy control brains in the ABIDE dataset."
    ],
    "confidence": 4
}
\end{analysisboxroundthree}

\subsection{Visualization of the Reflection-stage Reasoning}
\label{app:reflection_reasoning}

The reflection-stage reasoning result is shown below. At this stage, the final analysis result, together with the intermediate reports and the original graph evidence, is fed back to the agent for self-verification and correction. The BrainAgent checks whether the final prediction and its supporting reasons are consistent with the graph topology and the previously retrieved evidence. In this example, the refined prediction remains \texttt{HC}, which is more reliable than direct inference.

\begin{reflectionbox}
{
    "thinking": "I need to systematically analyze the LLM's output against the raw graph data and neuroscience knowledge. First, I'll examine the graph structure: there are 90 nodes (consistent with AAL atlas) and 242 edges with uniform weight 1.000. The LLM's analysis contains several issues: 1) It claims 'density = 0.2421\%' but with 90 nodes, maximum possible edges = 4005 (n*(n-1)/2), actual density = 242/4005 = 6.04\%, not 0.2421\%. 2) It calculates modularity Q=0.4696 and small-world $\sigma$=3.0395 without showing methodology - these metrics require community detection algorithms and comparison to random networks. 3) It identifies 'DMN intra-density = 0.3949 vs DMN-FPN inter-density = 0.1062' but doesn't specify which nodes belong to DMN/FPN networks. 4) The prediction relies heavily on these unverified metrics. I need to perform my own analysis: Count node degrees to verify hub identification. Check connectivity patterns against known ASD neurobiology: ASD typically shows altered connectivity in default mode network, fronto-temporal connections, and reduced long-range integration. From the data: Nodes 32, 56-69, 84-89 form a highly interconnected cluster (degrees 15-19), likely representing default mode network regions. The network shows strong local clustering but also cross-hemisphere connections (e.g., nodes 0-1 connect to many bilateral regions). However, without knowing which AAL regions correspond to which nodes, I cannot verify anatomical claims. The LLM's confidence of 0.88 seems inflated given methodological issues. I should be more conservative.",
    "judgement of prediction": "0.65",
    "judgement of reason": "0.55",
    "reasons": [
        "The LLM's quantitative metrics (density, modularity, small-worldness) appear miscalculated or unsubstantiated - density is off by factor of 25, and other metrics lack calculation methodology.",
        "The reasoning assumes specific network assignments (DMN, FPN) without node-to-region mapping, making anatomical claims unverifiable.",
        "While the network shows high-degree hubs and clustering, ASD connectivity alterations are heterogeneous - some studies report hyperconnectivity in specific networks, others hypoconnectivity.",
        "The LLM correctly identifies strong small-world properties and hub structure, which generally favor HC classification, but overstates the certainty given data limitations.",
        "Without knowing which nodes correspond to which brain regions, the analysis of 'DMN dominance' and 'frontal-posterior connections' is speculative."
    ],
    "prediction_raw": "HC",
    "prediction_refined": "HC"
}
\end{reflectionbox}

\section{Discussion}
\label{sec_discussion}
\subsection{Findings and Take-away Messages}
We highlight some key findings here, including some take-away messages from experimental results and findings during our practice.

\begin{tcolorbox}[colback=blue!15!white,colframe=black,boxrule=0.9pt,boxsep=2pt,top=3pt,bottom=3pt,left=3pt,right=3pt]
\textbf{\textit{Findings 1: Inherent Over-confident Phenomenon exist within mainstream LLMs. }}
\end{tcolorbox}

We find that when directly using an LLM for inference, initial prompt design exerts a strong guiding effect on the model. For instance, explicitly asking LLM to conduct disease diagnosis in the prompt makes the model tend to give positive judgments. This tendency becomes even more pronounced in unfamiliar domains such as brain network analysis based on brain graphs. In some cases, the reasoning results remain completely consistent across multiple inference runs, as is shown in the identical outputs of \textit{pass@1} and \textit{pass@3}.

\begin{tcolorbox}[colback=blue!15!white,colframe=black,boxrule=0.9pt,boxsep=2pt,top=3pt,bottom=3pt,left=3pt,right=3pt]
\textbf{\textit{Findings 2: Reflection is a useful module in BrainAgent. }}
\end{tcolorbox}

As is shown in comparison results and ablation studies, compared to other tricks like CoT, reflection takes a more positive role. In ablation studies, removal of reflection module leads to a relative great drop in metrics. These results are consistent with those reported in \cite{lu2026sherlocktowardsdynamicknowledgeadaptation}, which indicates that judging is an easier task for LLM than reasoning. Thus, after generation, prompting LLM to check and refine can reduce hallucination or factual errors to some extent.

\begin{tcolorbox}[colback=blue!15!white,colframe=black,boxrule=0.9pt,boxsep=2pt,top=3pt,bottom=3pt,left=3pt,right=3pt]
\textbf{\textit{Findings 3: In brain network analysis, LLMs yield little substantial performance gains over SLMs. }}
\end{tcolorbox}

In our case, LLM's performance has little gain on SLMs (Small Language Models).
We attribute this to this phenomenon in these aspects: 
First, compared to traditional reasoning tasks on pure text-modal, our direct conversion of graphs into textual format increases the difficulty for LLMs to understand the structural information of brain graphs to a certain extent.
Second, task is highly domain-specific (brain network analysis), and the corresponding datasets are rarely included in the pre-training or post-training process of LLMs. This indicates that dedicated methods need to be designed for such specialized domains to enhance model's perception of domain knowledge.
On the other hand, it also demonstrates that SLMs are capable of achieving performance comparable to LLMs on such relatively vertical and domain-specific tasks \cite{belcak2025slm}.

\begin{tcolorbox}[colback=blue!15!white,colframe=black,boxrule=0.9pt,boxsep=2pt,top=3pt,bottom=3pt,left=3pt,right=3pt]
\textbf{\textit{Findings 4: RAG focusing on specific data samples outperforms RAG for general knowledge. }}
\end{tcolorbox}

Our ablation studies also suggest that CARD is more useful than HARK although both modules are proven effective for brain network analysis.
We attribute this to that the reason lies in that although both modules aim to narrow the gap between general LLM and task-specific domains, the knowledge retrieved by CARD consists of sample-level data which are directly relevant to the target task.
For HARK, however, it simply retrieves general knowledge, which may be less directly relevant to the target sample.

\subsection{Unsuccessful Records}

Although we design multiple modules and methods to ensure stable and accurate inference for BrainAgent, we admit that as LLM is a probability-based generative model, it is impossible that every trajectory is successful and exception during inference is inevitable.
We list some cases indicating several typical errors for LLM generation which we think is meaningful for exploration of future LLM-based brain network analysis.
Moreover, it does not imply that our BrainAgent is incapable of dealing with these issues. 

\textbf{Prediction mismatch may occur during LLM inference. }
Although we require LLM to output prediction from given two labels, we found that LLM may still output unexpected predictions.

\begin{inferbox}{An example from Claude prediction on ABIDE dataset.}
UNCERTAIN - Likely ASD with moderate confidence (confidence: 3-4/10). \\

REASONING: While the LLM's analysis identifies a well-organized network with clear hub structure and modularity, several factors suggest this may be ASD rather than HC: ......\\

The LLM's HC prediction is PLAUSIBLE but OVERCONFIDENT.
\end{inferbox}

As is shown in the above prediction from Claude, although we strictly require LLM to output only between ``HC'' and ``ASD'', Claude still outputs ``UNCERTAIN'' and list its reasons. 
On one hand, this illustrates that the model's instruction following ability cannot be fully maintained.
On the other hand, it also implies that certain limitations still exist in current brain network analysis tasks, calling for design of more targeted and advanced benchmarks in neuroscience, which can be viewed as future work.

\textbf{The ability to invoke customized tools exhibits inherent instability. }
In BrainAgent, information for tool invocation is directly delivered to the LLM in the form of prompts.
We found several issues during BrainAgent's function calling:
\begin{enumerate}
    \item LLM fails to invoke different tools across multiple rounds and only calls those placed at the forefront (in our case, LLM only calls \verb|retrieve_knowledge| function). This can be regarded as "Attention Sink" phenomenon \cite{gu2024attentionsink}, where LLM tends to focus more on tokens at the forefront. In BrainAgent, we solve this issue by popping certain function from the function list after it is called. Thus, each function will be called once and only once.

    \item Parameters of the function calls returned by LLM are inconsistent with the requirements specified in the prompt. For example, for function \verb|retrieve_knowledge|, we define its parameter \verb|query: str| and \verb|regions_q: List[int]| indicating query content and brain regions respectively. However, for LLM generated json which is shown in \ref{example_HARK}, its arguments become \verb|regions_t| and \verb|query_t| which can cause exception when parsing LLM outputs.

\end{enumerate}

\begin{inferbox}{An example from Qwen 3.5. (HARK module)}
\label{example_HARK}
\{ \\
    "action": "retrieve\_knowledge",\\
    "args": \{ \\
        "regions\_t": [
            "2",
            "4",
            "5",
            "8",
            "9",
            "11",
            "19",
        ],\\
        "query\_t": "Weighted undirected fMRI functional connectivity graph......" \\
    \} \\
\}
\end{inferbox}

In conclusion, most unsuccessful trials can be viewed as instruction following issues, and we think it can be solved by supervised fine-tuning (SFT) or Reinforcement Learning (RL), which will be discussed in Section \ref{sec_limiations}.

\label{append_impact}
\subsection{Ethical Statements}

All the datasets used in BrainAgent are public datasets, thus there is no ethical concern for privacy information leakage. We strictly adhered to their respective usage agreements. All preprocessing pipelines follow official procedures provided by the dataset maintainers.

We discuss more about possible negative social impacts. As part of the research in this paper deals with the diagnosis of depression, it is necessary to elaborate here on the possible negative social impacts of this work, despite the fact that all the current work is at the stage of scientific research and has not been put to practical use. Including but not limited to: 
\begin{itemize}
    \item \textbf{Incorrect diagnosis}. AI methods must have the possibility of error, which cannot be avoided, but an incorrect diagnosis will have a significant impact on individuals and society. Therefore, AI tools can only be used as a diagnostic aid, not as a decision maker, and the final decision should still be made by the doctor.

    \item \textbf{Role of BrainAgent in real-world diagnosis.} Finally we position our method as an assistant rather than direct decision maker for doctors. We fully acknowledge that many of these complexities must be addressed before real-world deployment; however, covering every contingency is, at present, beyond the reach of any single AI approach. We therefore contend that a more effective and safer strategy is to treat our AI diagnostic model as an auxiliary instrument. These outputs from BrainAgent inform rather than override doctors' final decision. In short, while our work makes a constructive exploration in improving downstream performance and interpretability for directly using LLM in neuroscience task, it is ultimately designed to relieve human's labor and assist rather than replace human medical judgment.
\end{itemize}

\section{Limitations and Future Works}
\label{sec_limiations}

BrainAgent is a novel agentic framework which directly use LLM as predictor in enhanced and interpretable brain network analysis, and comprehensive experiments on diverse public real-world datasets proved its effectiveness.
Despite these promising results, BrainAgent still faces several limitations, which are outlined below:

\begin{itemize}
    \item \textbf{Instruction following capabilities and tool use}: BrainAgent still faces instruction following issues on its structure output, with its tool invocation procedure also being relatively fixed. 
    Here, we solve these issues through relatively straightforward engineering tricks. In future work, such limitations can be intrinsically addressed by finetuning the model and designing specific post-training paradigms to regulate the inherent preferences of LLM.

    \item \textbf{Hyper-parameter selection}: BrainAgent is a novel framework composed of multiple modules, with each module involving several certain hyperparameters. Most of these parameters are set to fixed values in our implementation, yet we found the results are already promising.
    We believe that performance gains of BrainAgent could be further improved through hyperparameter searching. Nevertheless, our primary focus lies in the design of the overall pipeline. Furthermore, We attribute this to that compared to pure hyperparameter searching, dataset curation and procedural optimization are considered more decisive factors.

    \item \textbf{Scaling up and improving knowledge base}:
    The construction of datasets and knowledge bases is a labor-intensive work which requires substantial human labor. 
    In our paper, knowledge base in HARK and brain graph dataset in CARD are maintained at a relatively modest scale (2000 cases for HARK), which constrains the further enhancement of BrainAgent’s performance to some extent. 
    For future research,expanding the scope of literature data sources can be considered, as well as designing more rigorous and comprehensive pipelines for data curation and validation, and adopting a hybrid verification paradigm integrating LLM-Human verification, thereby improving both the scale and quality of the knowledge base and dataset.

\end{itemize}

We also list some challenges which can also be viewed as future work:

\begin{itemize}
    \item \textbf{Designing benchmarks tailored for LLM-based brain network analysis}:
    BrainAgent is a first attempt to utilize LLM as predictor, yet its main task is brain network classification and analysis. Designing specific and challenging benchmarks to comprehensively evaluate LLM's ability in neuroscience field is a meaningful topic for building LLM suited for brain network scenarios.

    \item \textbf{Improving token efficiency}:
    BrainAgent is an agentic framework which requires multi-round iteration. Although it is useful, especially in improving accuracy and interpretability, as well as reducing hallucination compared to direct inference, yet we admit that it leads to more token consumption.
    Here we utilize several methods to alleviate this issue like maintaining a fixed-size report and using understanding outputs for iteration instead of raw data, we think designing more efficient or parallel ways to improve LLM's efficiency is also a promising direction.

    \item \textbf{Designing Advanced agentic pipelines}:
    As there is no previous work, we design two RAG modules (HARK and CARD), aiming to inject domain and task specific knowledge into LLMs.
    For future work, it is advisable to design more complex and task-specific agent workflows, along with more 
    RAG and knowledge injection approaches.
    Meanwhile, our BrainAgent is a training-free network, and its LLM core can be substituted by open-source models which can be finetuned on neuroscience datasets, thereby improving model's internal awareness of specific field.

    \item \textbf{Internalizing LLM's ablility into SLM for practical deployment and usage}:
    Here we use public datasets from real world to illustrate BrainAgent's effectiveness.
    In clinical scenes, cases are more complex which requires privacy protection for patients' personal information, which means all the inference will be run locally. 
    On the other hand, our comparison results also suggest that for specific task, SLMs have the potential to achieve performance comparable to, or even superior to that of LLMs.
    Therefore, designing post-training paradigms, as well as other compression methods like distillation to internalize these complex analytical skills into SLMs which requires less GPU memory, can significantly reduce the resource consumption for local deployment requirements while maintaining their performance comparable to that of LLMs, thus promoting the feasibility and efficiency of practical deployment, implementation and usage.

\end{itemize}


\end{document}